\documentclass[longauth]{aa}

\usepackage{graphicx}
\usepackage{color}
\usepackage{url}
\usepackage{multirow}

\usepackage{txfonts}
\begin{document} 

	\title{The eROSITA Final Equatorial-Depth Survey (eFEDS):}
    \subtitle{A multiwavelength view of WISE mid-infrared galaxies/active galactic nuclei}

   \author{	Yoshiki Toba \inst{1,2,3,4},
   			Teng Liu \inst{5},
			Tanya Urrutia \inst{6},
			Mara Salvato \inst{4},	 
			Junyao Li \inst{7,8,9},
			Yoshihiro Ueda \inst{1},
			Marcella Brusa \inst{10,11}, 
			Naomichi Yutani \inst{12},
			Keiichi Wada \inst{12,3,13},
			Atsushi J. Nishizawa \inst{14}, 
			Johannes Buchner \inst{4},
			Tohru Nagao \inst{3},
			Andrea Merloni \inst{4},	
			Masayuki Akiyama \inst{15},
			Riccardo Arcodia \inst{4},	
			Bau-Ching Hsieh \inst{2},
			Kohei Ichikawa \inst{16,15,4},
			Masatoshi Imanishi \inst{17,18},
			Kaiki T. Inoue \inst{19},	
			Toshihiro Kawaguchi \inst{20},
			Georg Lamer \inst{5},		
			Kirpal Nandra \inst{4},	
			John D. Silverman \inst{7,21},
			Yuichi Terashima \inst{3}
          }

   \institute{Department of Astronomy, Kyoto University, Kitashirakawa-Oiwake-cho, Sakyo-ku, Kyoto 606-8502, Japan\\
            \email{toba@kusastro.kyoto-u.ac.jp} 
            \and
			Academia Sinica Institute of Astronomy and Astrophysics, 11F of Astronomy-Mathematics Building, AS/NTU, No.1, Section 4, Roosevelt Road, Taipei 10617, Taiwan 
			\and
			Research Center for Space and Cosmic Evolution, Ehime University, 2-5 Bunkyo-cho, Matsuyama, Ehime 790-8577, Japan 
			\and			
			Department of Physics, Nara Women's University, Kitauoyanishi-machi, Nara, Nara 630-8506, Japan
			\and
			Max-Planck-Institut f\"ur Extraterrestrische Physik (MPE), Giessenbachstrasse 1, 85748 Garching bei M\"unchen, Germany	
		   	\and
			Leibniz-Institut f\"ur Astrophysik, Potsdam (AIP), An der Sternwarte 16, 14482 Potsdam, Germany
			\and
		   CAS Key Laboratory for Research in Galaxies and Cosmology, Department of Astronomy, University of Science and Technology of China, Hefei 230026, China
			\and
			Kavli Institute for the Physics and Mathematics of the Universe (WPI), The University of Tokyo, Kashiwa, Chiba 277-8583, Japan
			\and
			School of Astronomy and Space Science, University of Science and Technology of China, Hefei 230026, China	
			\and	
			Dipartimento di Fisica e Astronomia, Universit\`a di Bologna, via Gobetti 93/2, 40129 Bologna, Italy
			\and
			INAF- Osservatorio di Astrofisica e Scienza dello Spazio di Bologna, via Gobetti 93/3, 40129 Bologna, Italy
			\and
			Kagoshima University, Graduate School of Science and Engineering, Kagoshima 890-0065, Japan
			\and
			Hokkaido University, Faculty of Science, Sapporo 060-0810, Japan
			\and
			Institute for Advanced Research, Nagoya University, Furo-cho, Chikusa-ku, Nagoya, Aichi 464-8602,
Japan	
           	\and
           	Astronomical Institute, Tohoku University, 6-3 Aramaki, Aoba-ku, Sendai, Miyagi 980-8578, Japan
			\and
			Frontier Research Institute for Interdisciplinary Sciences, Tohoku University, Sendai 980-8578, Japan
			\and
			National Astronomical Observatory of Japan, National Institutes of Natural Sciences (NINS), 2-21-1 Osawa, Mitaka, Tokyo 181-8588, Japan
			\and
			Department of Astronomy, School of Science, The Graduate University for Advanced Studies, SOKENDAI, Mitaka, Tokyo 181-8588, Japan
           	\and
		   	Faculty of Science and Engineering, Kindai University, Higashi-Osaka, 577-8502, Japan
		   	\and
		   	Department of Economics, Management and Information Science, Onomichi City University, Hisayamada 1600-2, Onomichi, Hiroshima 722-8506, Japan
			\and
			Department of Astronomy, School of Science, The University of Tokyo, 7-3-1 Hongo, Bunkyo, Tokyo 113-0033, Japan
			}
\date{\today}

 
  \abstract
   {}
    {We investigate the physical properties---such as the stellar mass ($M_*$), star-formation rate (SFR), infrared (IR) luminosity ($L_{\rm IR}$), X-ray luminosity ($L_{\rm X}$), and hydrogen column density ($N_{\rm H}$)---of mid-IR (MIR) galaxies and active galactic nuclei (AGN) at $z < 4$ in the 140 deg$^2$  field observed by SRG/{\it eROSITA} using the Performance-and-Verification-Phase program named the {\it eROSITA} Final Equatorial Depth Survey (eFEDS).}
     {By cross-matching the {\it WISE} 22 $\mu$m (W4)-detected sample and the eFEDS X-ray point-source catalog, we find that 692 extragalactic objects are detected by {\it eROSITA}.
   We have compiled a multiwavelength dataset extending from X-ray to far-IR wavelengths. 
   We have also performed (i) an X-ray spectral analysis, (ii) spectral-energy-distribution (SED) fitting using {\tt X-CIGALE}, (iii) 2D image-decomposition analysis using Subaru Hyper Suprime-Cam (HSC) images, and (iv) optical spectral fitting with {\tt QSFit} to investigate the AGN and host-galaxy properties. 
   For 7,088 {\it WISE} 22 $\mu$m objects that are undetected by {\it eROSITA}, we have performed an X-ray stacking analysis to examine the typical physical properties of these X-ray faint and/or probably obscured objects.}
     {We find that (i) 82\% of the eFEDS--W4 sources are classified as X-ray AGN with $\log\,L_{\rm X} >$ 42 erg s$^{-1}$; (ii) 67\% and 24\% of the objects have $\log\,(L_{\rm IR}/L_\sun) > 12$ and 13, respectively; (iii) the relationship between $L_{\rm X}$ and the 6 $\mu$m luminosity is consistent with that reported in previous works; and (iv) the relationship between the Eddington ratio and $N_{\rm H}$ for the eFEDS--W4 sample and a comparison with a model prediction from a galaxy-merger simulation indicates that approximately 5.0\% of the eFEDS--W4 sources in our sample are likely to be in an AGN-feedback phase, in which strong radiation pressure from the AGN blows out the surrounding material from the nuclear region.}
     {Thanks to the wide area coverage of eFEDS, we have been able to constrain the ranges of the physical properties of the {\it WISE} 22 $\mu$m-selected sample of AGNs at $z < 4$, providing a benchmark for forthcoming studies on a complete census of MIR galaxies selected from the full-depth {\it eROSITA} all-sky survey.}
           

  	\titlerunning{WISE 22 $\mu$m-selected galaxies in the eFEDS}
	\authorrunning{Yoshiki Toba et al.}

   \maketitle
%

\section{Introduction}

Since infrared (IR) all-sky surveys were originally conducted with the {\it Infrared Astronomical Satellite} \citep[{\it IRAS};][]{Neugebauer} and the {\it AKARI} satellite \citep{Murakami} as well as deep IR observations with the {\it Infrared Space Observatory} \citep[{\it ISO};][]{Kessler}, IR galaxies have been established as an important population for understanding both galaxy formation and evolution and the co-evolution of galaxies and supermassive black holes (SMBHs) \citep[see e.g.,][and references therein]{Sanders,Goto,ChenX}.
The IR galaxies are mainly powered by star formation (SF), active galactic nuclei (AGN), or both, and these emissions contribute to the cosmic IR background \citep[e.g.,][]{Lagache}.

Following the abovementioned pioneering works, the {\it Spitzer} Space Telescope \citep{Werner} has shed light on mid-IR (MIR) objects in the high-$z$ universe \citep[see the review by][and references therein]{Soifer}.
More than 300,000 MIR galaxies have been found with flux densities at 24 $\mu$m ($f_{\rm 24}$) ranging down to several tens of $\mu$Jy, and their number counts, energy (AGN/SF) diagnostics, and contributions to the cosmic SF density have been investigated extensively \citep[e.g.,][]{Chary,Donley,Le,Magnelli}.
A relevant result of these studies on ``faint'' IR galaxies is that the number fraction of AGN increases with increasing MIR flux \citep[e.g.,][]{Brand,Treister} \citep[see also][]{Veilleux}.

The advent of the {\it Wide-field Infrared Survey Explorer} \citep[{\it WISE};][]{Wright} has provided an avenue for detecting an enormous number of MIR AGN \citep[see e.g.,][]{Assef}.
In particular, MIR-bright ($f_{\rm 22}$ > a few mJy) but optically faint WISE objects [which are often called dust-obscured galaxies (DOGs)\footnote{The original definition of DOGs was $f_{\rm 24} > 0.3$ mJy and $R-$[24] $>$ 14, where $R$ and [24] represent Vega magnitudes in the $R$-band and at 24 $\mu$m, respectively \citep[see][for more detail]{Dey_08}.}], have been reported to harbor a strong AGN surrounded by a large amount of dust \citep[e.g.,][]{Eisenhardt,Wu,Toba_15,Toba_16,Toba_17a,Noboriguchi,Yutani}.
Some dusty AGN are located at $1 < z < 4$, where the cosmic star formation rate (SFR) density and BH mass-accretion rate density reach a maximum \citep[e.g.,][]{Madau,Ueda}. Thus, these objects may be a crucial population for unveiling the growth history of SMBHs and their host masses in the dusty universe.

From an X-ray study point of view, a large fraction of (dusty) IR galaxies was not detected at X-ray wavelengths \citep[see the review by][]{Hickox}.
According to deep X-ray observations with {\it Chandra}, a fraction of the {\it Spitzer}-detected IR-faint AGN may be expected to have a hydrogen column density as large as $N_{\rm H} = 10^{23}$ cm$^{-2}$. Some are even Compton-thick (CT) AGN, with $N_{\rm H}$ $> 1.5 \times 10^{24}$ cm$^{-2}$, although the fraction of CT-AGN has quite a large variation---from 10\% to 90\%--- possibly caused by differences in sample selection and analyses \citep{Fiore_08,Lanzuisi,Treister_09,Corral}.
In particular, the fraction of CT-AGN among DOGs may increase with increasing IR luminosity \citep{Fiore_08}.
Recently, \cite{Carroll} reported that about 62\% of IR-bright {\it WISE} AGN are not detected in X-rays.
Using the intrinsic-to-observed X-ray luminosity ratio, they found that those AGN have high $N_{\rm H}$, approaching the CT-AGN regime, which is supported by survival analysis and X-ray stacking analysis.
\cite{Toba_20a} indeed discovered a CT-AGN using deep {\it NuSTAR} observations of an IR-bright DOG.
However, studies of MIR galaxies at $z < 4$ with moderately deep X-ray surveys are still based on a limited survey area, $<$ 50 deg$^2$ \citep[e.g.,][and references therein]{LaMassa,Mountrichas}, which prevents us from obtaining a complete picture of the X-ray properties of MIR galaxies, such as their X-ray luminosity, over a wide parameter range.

The {\it extended ROentgen Survey with an Imaging Telescope Array} \citep[{\it eROSITA};][]{Merloni,Merloni_20,Predehl} provides a valuable probe for determining the X-ray properties of galaxies.
{\it eROSITA} is the primary instrument on the Spectrum-Roentgen-Gamma (SRG) mission \citep{Sunyaev}, which was successfully launched on July 13, 2019 \citep[see also][]{Pavlinsky}.
The combination of {\it WISE} and {\it eROSITA} provides a complete census of MIR galaxies over a wide range of X-ray luminosities.
Because of the all-sky surveys, even if {\it WISE} MIR sources are not detected by {\it eROSITA}, an X-ray stacking analysis can still yield estimates of the typical X-ray properties of this possibly heavily obscured population, as successfully achieved with X-ray deep fields \citep[e.g.,][]{Daddi,Eckart}.

In this paper, we report a multiwavelength analysis of MIR galaxies at $z < 4$ in the GAMA-09 field observed by {\it eROSITA} using the Performance-and-Verification-Phase program named the ``{\it eROSITA} Final Equatorial Depth Survey \citep[eFEDS:][]{Brunner}.''
The eFEDS main X-ray catalog contains 27,369 X-ray point sources detected over an area of 140 deg$^2$ in a single broad band, with a 5$\sigma$ sensitivity of $f_{\rm 0.3--2.3~keV} \sim 9 \times 10^{-15}$ erg s$^{-1}$ cm$^{-2}$ \citep{Liu,Salvato_21}, which reveals X-ray properties of more than 500 MIR-detected AGN. 
The structure of this paper is as follows.
Sect.~\ref{DA} describes sample selection, the multiwavelength dataset, and the methodology of (i) X-ray spectral fitting, (ii) spectral-energy-distribution (SED) modeling, (iii) 2D image decomposition, and (iv) X-ray stacking analysis.
The results of these analyses are summarized in Sect.~\ref{R}.
In Sect.~\ref{D}, we discuss the physical properties of AGN and their hosts and characterize the {\it eROSITA}-detected and -undetected {\it WISE} MIR galaxies by comparison with a model prediction from a galaxy-merger simulation.
Finally, our conclusions and a summary are given in Sect.~\ref{Sum}.
Throughout this paper, the adopted cosmology is a flat universe with $H_0$ = 70 km s$^{-1}$ Mpc$^{-1}$, $\Omega_M$ = 0.3, and $\Omega_{\Lambda}$ = 0.7, which are the same as those adopted in \cite{Liu} and \cite{Salvato_21}.
We assume the initial mass function (IMF) of \cite{Chabrier}.
Unless otherwise noted, all magnitudes refer to the AB system.

\section{Data and analysis}
\label{DA}

\subsection{Sample selection}
\label{SS}

\begin{figure}
\centering
\includegraphics[width=0.45\textwidth]{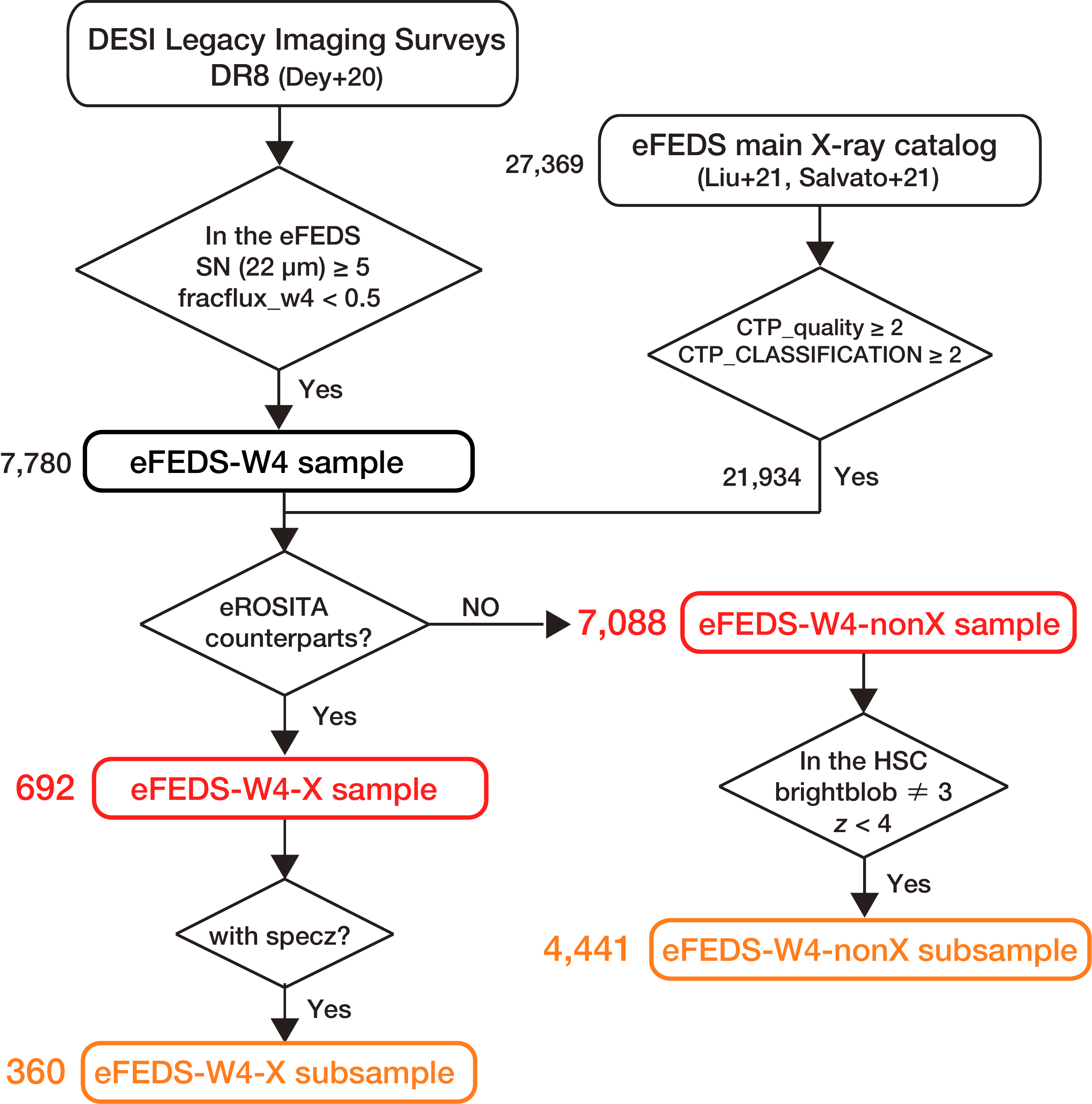}
\caption{Flowchart of the sample-selection process.}
\label{fig_SS}
\end{figure}

Figure \ref{fig_SS} shows a flowchart of our sample-selection process.
Our sample is drawn from the DESI Legacy Imaging Surveys Data Release (DR) 8\footnote{\url{https://www.legacysurvey.org/dr8/}} \citep[LS8:][]{Dey_19}.
LS8 provides optical photometry ($g$, $r$, and $z$) taken from the Dark Energy Camera Legacy Survey (DECaLS) and {\it WISE} forced photometry from imaging through {\it NEOWISE-Reactivation} \citep[{\it NEOWISE-R};][]{Mainzer} that is measured in the {\it unWISE} maps \citep{Lang,Lang16} at the locations of the optical sources.
{\it WISE} provides four-band MIR photometry at 3.4 $\mu$m (W1), 4.6 $\mu$m (W2), 12 $\mu$m (W3), and 22 $\mu$m (W4). 
In addition, LS8 provides photometry, parallaxes, and proper motions from {\it Gaia} DR2 \citep{Gaia}.

We first prepared a {\it WISE} 22 $\mu$m (W4)-detected sample with a signal-to-noise ratio (SN) of $f_{\rm 22} > 5$ from LS8.
We then excluded the objects for which $f_{\rm 22}$ was severely affected by contributions from their neighborhoods by adopting {\tt fracflux\_w4}\footnote{\url{https://www.legacysurvey.org/dr8/catalogs/}} $< 0.5$, which yielded 7,780 objects (hereafter referred to as the eFEDS--W4 sample) in the eFEDS footprint\footnote{We visually confirmed that there were no spurious detections at 22 $\mu$m based on W4 images.}.
Next, we connected the eFEDS--W4 sample and the eFEDS main X-ray catalog with optical-to-MIR counterparts \citep{Salvato_21} through a unique ID for objects in LS8.
The counterparts were assigned after comparing the results of associations obtained using two independent methods: a Bayesian-statistics-based algorithm ({\tt NWAY}\footnote{\url{https://github.com/JohannesBuchner/nway}}; \citealt{Salvato}) and a maximum-likelihood method \citep{Sutherland} ({\tt astromatch}\footnote{\url{https://github.com/ruizca/astromatch}}; \citealt{Ruiz}).
Both of the methods used LS8; in particular, they were trained on the LS8 properties of a sample of more than 23,000 X-ray sources with secure counterparts \citep[see ][for more details]{Salvato_21}.
Based on the comparison of the two methods, each object was set to a counterpart reliability flag ({\tt CTP\_quality}).
\cite{Salvato_21} also provided a classification of counterparts (i.e., galactic or extragalactic source) based on optical spectra (if available), {\it Gaia} parallaxes, and/or multicolor information, which can be referred to as a classification flag ({\tt CTP\_CLASSIFICATION}).

\begin{figure}
\centering
\includegraphics[width=0.45\textwidth]{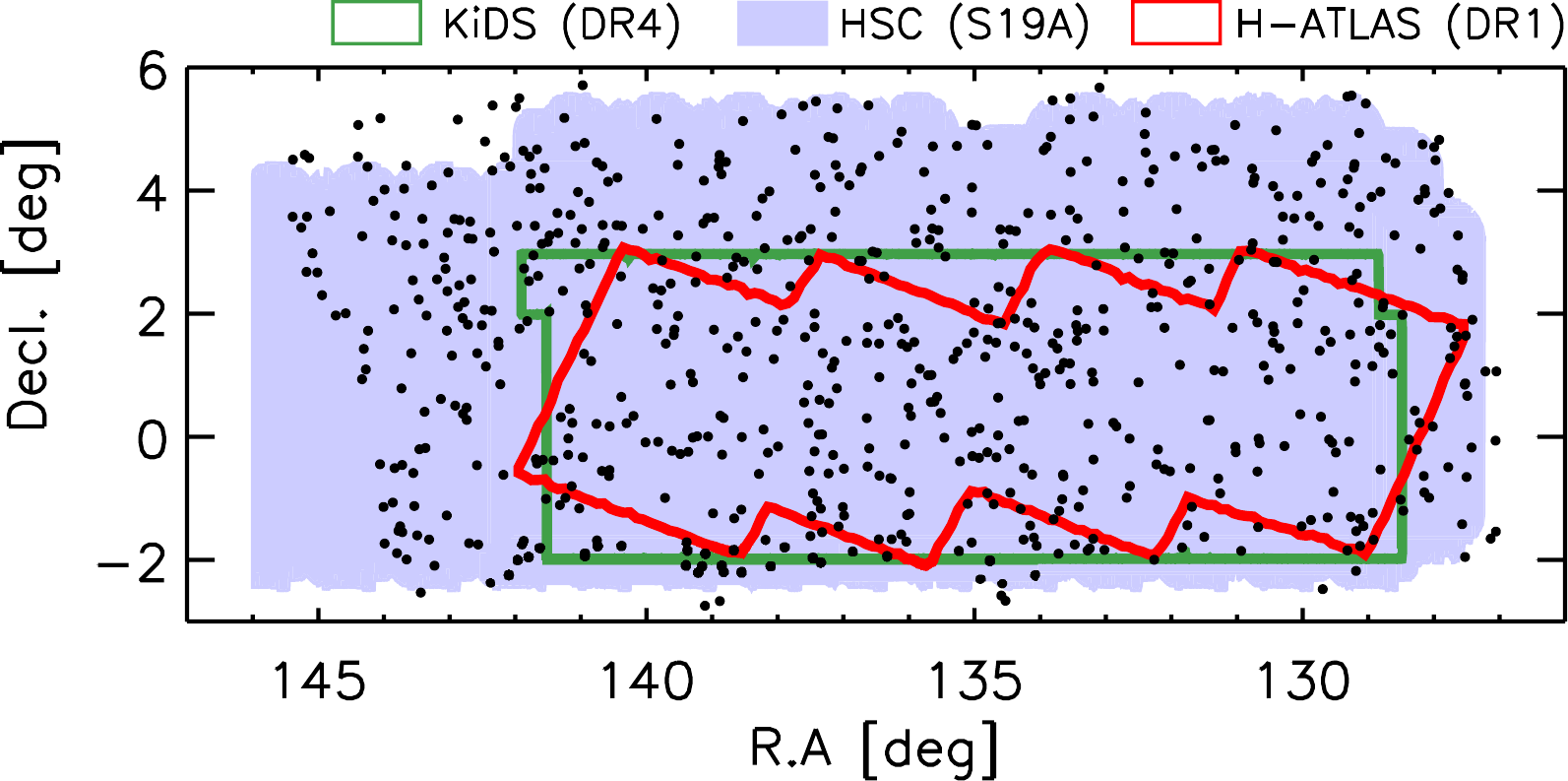}
\caption{Sky distribution of the 692 eFEDS--W4-X sample (black points) in the eFEDS catalog. The blue-shaded region represents the survey footprint of the HSC-SSP S19A, while the green and red outlines represent those of KiDS DR4 and H-ATLAS DR1, respectively.}
\label{fig_region}
\end{figure}

Before connecting the catalog, we reduced the eFEDS main X-ray sample to 21,934 extragalactic sources that had reliable optical counterparts by adopting {\tt CTP\_quality} $\geq$ 2 and {\tt CTP\_CLASSIFICATION} $\geq$ 2 \citep[see ][for more details]{Liu,Salvato_21}.
This left 692 objects with {\it eROSITA} counterparts (hereafter referred to as the eFEDS--W4-X sample); the corresponding sky distribution is presented in Fig.~\ref{fig_region}.
For the 7,088 objects that were not detected by {\it eROSITA} (hereafter referred to as the eFEDS--W4-nonX sample), we performed the stacking analysis described in Sect.~\ref{s_st}.
We then compiled redshifts and multiwavelength photometric data\footnote{\cite{Salvato_21} provided multiwavelength photometric data up to MIR for X-ray sources in the eFEDS main X-ray catalog. Nevertheless, we compiled data ourselves up to the far-IR (FIR) to perform SED analysis for the eFEDS--W4 sample regardless of {\it eROSITA} detections using identical photometry with UV--FIR (see Sect.~\ref{s_SED}).} for the eFEDS--W4 sample.

\subsection{Spectroscopic redshifts}
\label{s_photoz}

Spectroscopic redshifts ($z_{\rm spec}$\footnote{$z_{\rm spec}$ used in this work were obtained from the report by \cite{Salvato_21} except for a proprietary $z_{\rm spec}$, which was obtained through the SPectroscopic IDentification of ERosita Sources (SPIDERS)/SDSS-IV follow-up campaign (Merloni et al., in prep.).})  were compiled from the Two-Micron All-Sky Survey \citep[2MASS;][]{Skrutskie} Redshift Survey \citep[2MRS;][]{Huchra}, LAMOST DR4 \citep{Cui}, Sloan Digital Sky Survey \citep[SDSS;][]{York} DR16 \citep{Ahumada}, the 2dF-SDSS LRG and QSO Survey \citep[2SLAQ;][]{Croom}, Galaxy and Mass Assembly \citep[GAMA;][]{Driver} DR3 \citep{Baldry}, 6dF Galaxy Survey \citep[6dFGS;][]{Jones_04} DR3 \citep{Jones_09}, and WiggleZ Dark Energy Survey project DR1 \citep{Drinkwater}.

\begin{figure}
\centering
\includegraphics[width=0.45\textwidth]{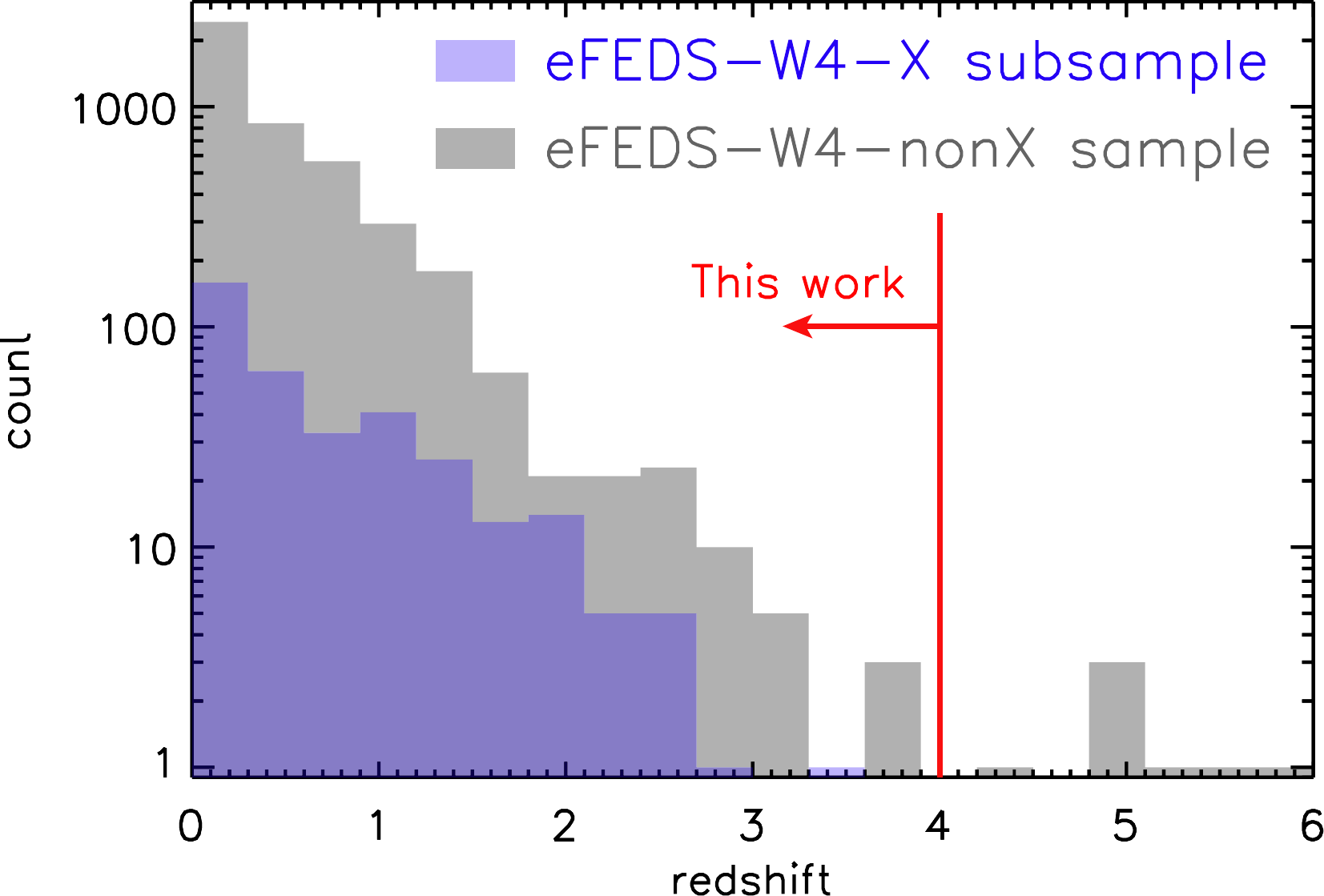}
\caption{Redshift distribution of the eFEDS--W4-X subsample (blue) and the eFEDS--W4-non X sample (gray). Objects at $z < 4$ are used for the multiwavelength analysis.}
\label{fig_zhist}
\end{figure}

In this work, we focused on objects whose redshifts were spectroscopically confirmed to derive reliable physical quantities and draw an accurate conclusion.
Therefore, we only considered 360 objects with $z_{\rm spec}$ (hereafter referred to the eFEDS--W4-X subsample) in Sects.~\ref{R} and \ref{D}.
The redshift distribution of the eFEDS--W4-X subsample is presented in Fig.~\ref{fig_zhist}.
The eFEDS--W4-X sources distribute the redshifts up to $z_{\rm spec} \sim 3.4$ with a mean redshift value of 0.63.

\subsection{Multiwavelength dataset}
\label{s_multi}
\subsubsection{UV data}
\label{s_UV}

Far-UV (FUV) and near-UV (NUV) data were taken from the {\it Galaxy Evolution Explorer} \citep[{\it GALEX};][]{Martin}.
We used the revised catalog\footnote{\url{http://dolomiti.pha.jhu.edu/uvsky/GUVcat/GUVcat_AIS.html}} from the {\it GALEX} All-Sky Survey (AIS) \citep[GUVcat\_AIS;][]{Bianchi}, which contains 82,992,086 sources with 5$\sigma$ limiting magnitudes of 19.9 in the FUV and 20.8 in the NUV.
Before cross-matching, we extracted 80,315,723 sources with (i) {\tt GRANK} $\leq$ 1 and (ii) {\tt fuv\_artifact} = 0 and {\tt fuv\_flags} = 0 or {\tt nuv\_artifact} = 0 and {\tt nuv\_flags} = 0, where {\tt GRANK}, {\tt f/nuv\_artifact}, and {\tt f/nuv\_flags} are the primary-source, artifact, and extraction flags, respectively. This eliminated spurious and duplicate sources \citep[see Sect.~6.2 and Appendix A in][for more detail]{Bianchi}.

\subsubsection{$u$-band data}
\label{s_u}

The $u$-band data were taken from the SDSS and the Kilo-Degree Survey \citep[KiDS;][]{de_Jong}.
We used the SDSS {\tt PhotoPrimary} table in DR16 \citep{Ahumada} and KiDS DR4\footnote{A small bug was reported by the KiDS team, which induces incorrect photometry in some tiles; for these tiles, the data in DR4.1 should be replaced. However, we confirmed that the eFEDS sources were not located in the problematic tiles; therefore, we used the original DR4 catalog (see \url{http://kids.strw.leidenuniv.nl/DR4/index.php}).} \citep{Kuijken}, which contain 469,053,874 and 100,350,871 sources, respectively.
The 5$\sigma$ limiting $u$-band magnitudes of the SDSS and KiDS DR4 are approximately 22.0 and 24.2, respectively.
Because KiDS DR4 does not cover the entire region of the eFEDS (see Fig. \ref{fig_region}), we used the SDSS $u$-band data for objects outside the KiDS footprint.
Before cross-matching, we extracted sources with {\tt FLAG\_GAAP\_u} = 0 to ensure reliable $u$-band fluxes for the KiDS sources \citep[see][]{Kuijken} (see also Sect. \ref{s_NIR}).

\subsubsection{Optical data}
\label{s_op}

In addition to DECaLS/LS8 ($g$, $r$, and $z$), we used KiDS DR4 ($g$, $r$, $i$, $z$, and $y$), where objects with {\tt FLAG\_GAAP\_g/r/i/z/y} = 0 were extracted in each band for cross-matching.
Additionally, we utilized deep optical data obtained from the Hyper Suprime-Cam \citep[HSC;][]{Miyazaki} Subaru Strategic Program survey \citep[HSC-SSP;][]{Aihara18a} \citep[see also][]{Furusawa,Kawanomoto,Komiyama}.
The HSC--SSP is an ongoing optical imaging survey with five broadband filters ($g$-, $r$-, $i$-, $z$-, and $y$-band) and four narrowband filters \citep[see][]{Aihara18a,Bosch,Coupon,Huang}.
In this work, we used S19A wide-layer data obtained from March 2014 to April 2019, for which the survey footprint mostly overlaps with eFEDS (see Fig.~\ref{fig_region}).
The HSC catalog provides forced photometry for the $g$-, $r$-, $i$-, $z$-, and $y$-bands with 5$\sigma$ limiting magnitudes of 26.8, 26.4, 26.4, 25.5, and 24.7, respectively \citep{Aihara18b,Aihara19}. These are approximately 15 times deeper than those from LS8.
The typical seeing is approximately 0$\farcs$6 in the $i$-band, and the astrometric uncertainty is approximately 40 mas in rms.

We first narrowed down the sample to unique HSC objects with {\tt isprimary} = True.
We then adopted five flags for each band: (i) {\tt sdsscentroid\_flag} = False (i.e., an object has a clean measurement of the centroid position), (ii) {\tt pixelflags\_edge} = False (i.e., an object is not at the edge of a CCD or a co-added patch), (iii) {\tt pixelflags\_saturatedcenter} = False (i.e., the central 3$\times$3 pixels of an object are not saturated, (iv) {\tt pixelflags\_crcenter} = False (i.e., the central 3$\times$3 pixels of an object are not affected by cosmic rays, and (v) {\tt pixelflags\_bad} = False (i.e., none of the pixels in the footprint of an object is labeled as bad).
This left 422,260,020 HSC sources for cross-matching. 

We employed clean optical data as much as possible for the SED fitting regardless of overlap of the survey footprints for LS8, KiDS, and HSC (see Sect.~\ref{s_SED}).

\subsubsection{Near-IR data}
\label{s_NIR}

We compiled near-IR (NIR) data from the VISTA Kilo-degree Infrared Galaxy Survey \citep[VIKING;][]{Arnaboldi,Edge}.
The VIKING data were incorporated into KiDS DR4, for which aperture-matched optical--NIR forced photometry was available.
We used $J$-, $H$-, and $K_{\rm S}$-band data with 5$\sigma$ limiting magnitudes of approximately 23 in $J$ and 22 in $K_{\rm S}$ \citep{Kuijken}.
Before cross-matching, we extracted sources with {\tt FLAG\_GAAP\_J} = 0, {\tt FLAG\_GAAP\_H} = 0, or {\tt FLAG\_GAAP\_Ks} = 0 to ensure reliable photometry \citep[see][]{Kuijken}.

Because the KiDS--VIKING catalog (KiDS DR4) partially covers the eFEDS footprint (see Fig. \ref{fig_region}), we also used the UKIRT Infrared Deep Sky Survey \citep[UKIDSS;][]{Lawrence} data.
We utilized the UKIDSS Large Area Survey DR11plus, which we obtained through the WSA--WFCAM Science Archive\footnote{\url{http://wsa.roe.ac.uk/index.html}}, which contains 88,298,646 sources.
The limiting magnitudes of UKIDSS are 20.2, 19.6, 18.8, and 18.2 Vega mag in the $Y$-, $J$-, $H$-, and $K$-bands, respectively.
The UKIDSS catalog contains the Vega magnitudes for each source, and we converted them into AB magnitudes using offset values $\Delta m$ ($m_{\rm AB} = m_{\rm Vega} + \Delta m$) for the $Y$-, $J$-, $H$-, and $K$-bands of 0.634, 0.938, 1.379, and 1.900, respectively, according to \cite{Hewett}.
Before cross-matching, we extracted 77,276,542 objects with ({\tt priOrSec} $\leq$ 0 or = {\tt frameSetID}) and ({\tt jpperrbits} $<$ 256 or {\tt hpperrbits} $<$ 256 or {\tt kpperrbits} $<$ 256) to ensure clean photometry for uniquely detected objects in the same manner as in \cite{Toba_15,Toba_19b}.
If an object lay outside the KiDS--VIKING footprint, its NIR flux densities were taken from UKIDSS; otherwise, we always referred to the KiDS--VIKING NIR data.

\subsubsection{Far-IR data}
\label{s_FIR}

We obtained the FIR data from a project of the {\it Herschel} Space Observatory \citep{Pilbratt} Astrophysical Terahertz Large Area Survey \citep[H-ATLAS;][]{Eales,Bourne}.
This survey provides flux densities at 100 and 160 $\mu$m---obtained using a photoconductor array camera and spectrometer \citep[PACS;][]{Poglitsch}---and at 250, 350, and 500 $\mu$m---obtained using the Spectral and Photometric Imaging Receiver \citep[SPIRE;][]{Griffin}.
We used H-ATLAS DR1 \citep{Valiante}, which contains 120,230 sources in the GAMA fields.
The 1$\sigma$ noise levels for source detection (which include both confusion and instrumental noise) are 44, 49, 7.4, 9.4, and 10.2 mJy at 100, 160, 250, 350, and 500 $\mu$m, respectively \citep{Valiante}.

Because H-ATLAS partially overlaps the eFEDS region (see Fig. \ref{fig_region}), we also utilized all-sky data from the {\it AKARI} FIR Surveyor \citep[FIS;][]{Kawada} Bright Source Catalog version 2.0 \citep{Yamamura}, which includes 918,054 FIR sources.
This catalog provides the positions and flux densities in four FIR wavelengths centered at 65, 90, 140, and 160 $\mu$m, with 5$\sigma$ sensitivities in each band of approximately 2.4, 0.55, 1.4, and 6.3 Jy, respectively; these are the deepest FIR all-sky data.
Following \cite{Toba_17b}, we selected 501,444 sources with high detection reliability ({\tt GRADE} = 3) for cross-matching, that is, sources that were detected in at least two wavelength bands or in four or more scans in one wavelength band.

\subsubsection{Cross identification of multi-band catalogs}
\label{s_CM}

We cross-identified the {\it GALEX}, KiDS--VIKING, SDSS, HSC, UKIDSS, {\it AKARI}, and H-ATLAS catalogs with the eFEDS--W4-X sample, always using the LS8 coordinates of the eFEDS--W4-X sample, which we refer to as R.A. and Decl.
Using a search radius of 1$\arcsec$ for KiDS--VIKING, SDSS, and HSC, 3$\arcsec$ for {\it GALEX}, 10$\arcsec$ for H-ATLAS, and 20$\arcsec$ for {\it AKARI} in the same manner as in \cite{Toba_17b,Toba_19a,Toba_19b}, we cross-identified 382 (55.2\%), 321 (46.4\%), 635 (91.8\%), 655 (94.7\%), 681 (98.4\%), 22 (3.2\%), and 119 (17.2\%) objects with those from {\it GALEX}, KiDS--VIKING, SDSS, HSC, UKIDSS, {\it AKARI}, and H-ATLAS, respectively.
We note that 3/382 (0.8\%), 33/635 (5.2\%), and 6/681 (0.9\%) objects had two candidates within the search radius as counterparts for the {\it GALEX}, HSC, and UKIDSS sources, respectively.
For cross-matching with other catalogs (SDSS, KiDS-VIKING, {\it AKARI}, and H-ATLAS), we obtained one-to-one identification. 
We found that the multiple fractions of HSC sources were relatively large, possibly due to the high number density of HSC sources.
Therefore, we estimated the chance coincidence of cross-matching with the HSC catalog by generating a mock catalog with random positions in the same manner as in \cite{Toba_19b}, where the source position in the HSC catalog was shifted from the original position to $\pm$1$\degr$ or $\pm$2$\degr$ along the right assignation direction.
Consequently, the chance coincidence was estimated to be $<0.2$\%.
We chose the nearest object as the counterpart for this case.

\subsection{X-ray spectral analysis}
\label{Xana}

\cite{Liu} analyzed the X-ray spectra of all eFEDS X-ray sources (see also Nandra et al., in prep.), including those of our eFEDS--W4-X sample.
The methodology is summarized below.
We extracted the X-ray spectra using \texttt{srctool} v1.63 of the {\it eROSITA} Science Analysis Software System \citep[eSASS;][]{Brunner}.
We used multiple models to fit the spectra of the AGN, and 
the baseline model was an absorbed power-law.
To model bright sources with a soft excess, we added an additional power-law component to the baseline model, creating a ``double-power-law'' model. 
To model faint sources for which we could not constrain the spectral shape parameters of the ``single-power-law'' model, we fixed the power-law slope $\Gamma$ and/or the absorbing column density N$_{\rm H}$ at typical values for the sample ($\Gamma=2.0$ and N$_{\rm H}=0$).

We employed the Bayesian method BXA \citep[][]{Buchner2014,Buchner2019} to derive constraints on the spectral parameters.
For each parameter, we measured the median and 1$\sigma$ confidence intervals from the posterior distribution.
In the spectral analysis, we always modeled the galactic absorption using N$_{\rm H}$, as measured by the HI4PI \citep{HI4PI}, and adopted the photoionization cross-sections provided in \cite{Verner} and the abundances provided in \cite{Wilms}.

Based on a detailed analysis of the parameter posterior distributions, \cite{Liu} provided an N$_{\rm H}$ measurement flag for each source, dividing the sources into four classes: 1) \texttt{uninformative}, 2) \texttt{unobscured}, 3) \texttt{mildly-measured}, and 4) \texttt{well-measured}.
Following \cite{Liu}, we considered \texttt{unobscured} sources with a median $\log\, N_{\rm H} < 21.5$ cm$^{-2}$ as X-ray unobscured (or type 1) AGN and \texttt{mildly-measured} or \texttt{well-measured} sources with a median $\log\,N_{\rm H} \geq 21.5$ cm$^{-2}$ as X-ray obscured (or type 2) AGN.
For each source, based on the parameter constraints of the data, we chose the most suitable model to estimate the X-ray luminosities. 
Generally, we adopted the flexible double-power-law model for high-quality data and the single-power-law model with one or more parameters fixed for low-quality data. 
However, in the case of X-ray type 2 AGN, we adopted the single-power-law model instead of the flexible double-power-law model to avoid strong degeneracy between the soft excess and the absorption.
A full description of the X-ray data reduction and analysis is provided in \cite{Liu}.

\begin{table}
\caption{Parameter ranges used in the SED fitting with {\tt X-CIGALE}.}
\label{Param}
\centering
\begin{tabular}{lc}
\hline \hline
Parameter & Value\\
\hline
\multicolumn2c{Double exp. SFH}\\
\hline
$\tau_{\rm main}$ [Myr] & 1,000, 2,000, 4,000, 6,000 \\
$\tau_{\rm burst}$ [Myr] & 3, 80 \\
$f_{\rm burst}$ & 0.01, 0.1 \\
age [Myr] & 500, 2,000, 4,000, 6,000 \\
\hline
\multicolumn{2}{c}{SSP \citep{Bruzual}}\\
\hline
IMF				&	\cite{Chabrier} \\
Metallicity		&	0.02  \\
\hline
\multicolumn2c{Nebular emission \citep{Inoue}}\\
\hline
$\log\, U$	&	$-2.0$	\\
\hline
\multicolumn2c{Dust attenuation \citep{Calzetti,Leitherer}}\\
\hline
$E(B-V)_{\rm lines}$ &  0.05, 0.1, 0.2, 0.4, 0.6, 0.8, 1.0, 2.0 \\
\hline
\multicolumn{2}{c}{AGN emission \citep{Stalevski12,Stalevski}}\\
\hline
$\tau_{\rm 9.7}$ 			&  	3, 7 		\\
$p$							&	0.5  	\\
$q$							&	0.5 	\\
$\Delta$ [$\degr$]			&	40				\\
$R_{\rm max}/R_{\rm min}$ 	& 	30 		\\
$\theta$ [$\degr$]			&	30, 50, 70		\\
$f_{\rm AGN}$ 				& 	0.1, 0.2, 0.3, 0.4, 0.5, 0.6, 0.7, 0.8, 0.9	\\
\hline
\multicolumn{2}{c}{Dust emission \citep{Dale}}\\
\hline
$\alpha$ & 0.0625, 2.0000, 4.0000 \\
\hline
\multicolumn{2}{c}{X-ray emission \citep{Yang}}\\
\hline
AGN photon index	&	1.5, 2.0, 2.5	\\
$|\Delta\,\alpha_{\rm OX}|_{\rm max}$	&	0.2	\\
LMXB photon index	&	1.56	\\
HMXB photon index	&	2.0		\\
\hline
\end{tabular}
\end{table}

\subsection{SED fitting with {\tt X-CIGALE}}
\label{s_SED}

To derive the physical properties of our sample, we performed SED fitting in which we employed a new version of the Code for Investigating GALaxy Emission \citep[{\tt CIGALE}; ][]{Burgarella,Noll,Boquien}, known as {\tt X-CIGALE} \footnote{\url{https://gitlab.lam.fr/gyang/cigale/tree/xray}} \citep{Yang}.
{\tt X-CIGALE} implements an X-ray module that enables fitting from X-ray to FIR wavelengths in a self-consistent framework by considering the energy balance between the UV/optical and IR wavelengths.
This code can handle many parameters, such as the star formation history (SFH), single stellar population (SSP), attenuation law, AGN emission, dust emission, and radio synchrotron emission \cite[see e.g.,][]{Boquien14,Boquien16,Buat,LoFaro,Toba_19b,Toba_20c}. 
The parameter ranges used in the SED fitting are listed in Table \ref{Param}.

\begin{figure*}
\centering
\includegraphics[width=0.7\textwidth]{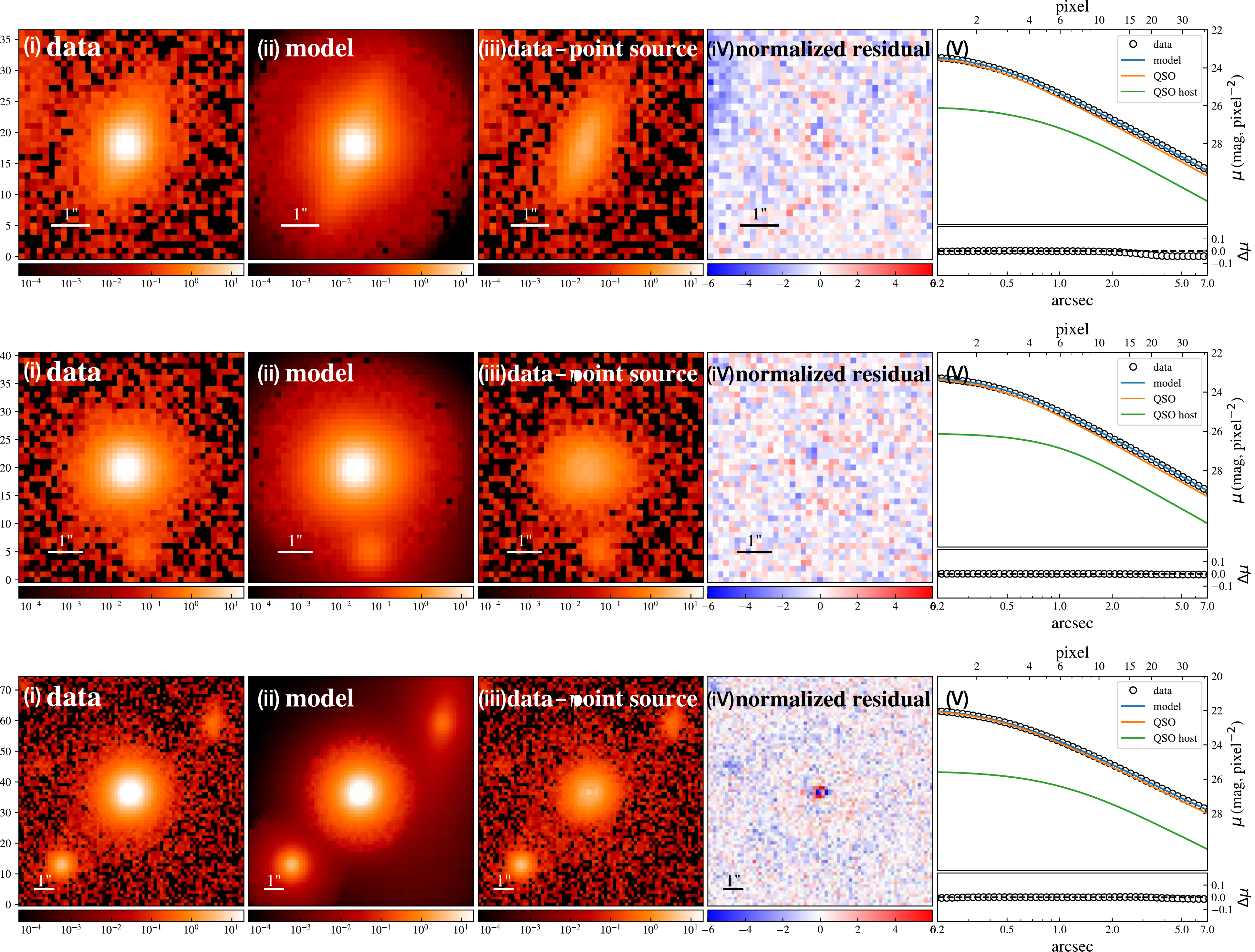}
\caption{Examples of quasar-host decomposition based on the HSC $i$-band image. From left to right: (i) the observed HSC $i$-band image, (ii) the best-fit point source + host galaxy model, (iii) the data minus the point source model (the pure-galaxy image), (iv) fitting residuals divided by the variance map, and (v) one-dimensional surface brightness profiles (top) and the corresponding residuals (bottom).}
\label{fig_decomp}
\end{figure*}

We assumed a SFH comprising two exponentially decreasing SFRs with different $e$-folding times \citep[e.g.,][]{Ciesla_15,Ciesla_16}.
We chose the SSP model of \cite{Bruzual}, assumed the IMF of \cite{Chabrier}, and used the standard nebular emission model included in {\tt X-CIGALE} \citep[see ][]{Inoue}.

For attenuation by the dust associated with the host galaxy, we utilized a starburst attenuation curve  \citep{Calzetti,Leitherer} in which we parameterized the color excess of the nebular emission lines [$E(B-V)_{\rm lines}$], extended with the \cite{Leitherer} curve.
The color excess of stars, $E(B-V)_{*}$, can be converted from $E(B-V)_{\rm lines}$ by assuming a simple reduction factor ($f_{\rm EBV}$ = $\frac{E(B-V)_{*}}{E(B-V)_{\rm lines}}$) = 0.44 \citep{Calzetti_97}.

We modeled the IR emission from dust (reprocessed from the absorbed UV/optical stellar emission) using the dust templates provided by \cite{Dale}: d$M_{\rm d}$ ($U$) $\propto U^{-\alpha}$d$U$, where $M_{\rm d}$ is the dust mass, $U$ is the intensity of the radiation field, and $\alpha$ is a dimensionless parameter.

We modeled the AGN emission using {\tt SKIRTOR} \citep{Baes,Camps,Stalevski12,Stalevski}, a two-phase medium torus model with high-density clumps embedded in a low-density component, with seven parameterized quantities \citep[see][for details]{Stalevski,Yang}.
 As parameters, we used the torus optical depth at 9.7 $\mu$m ($\tau_{\rm 9.7}$), the torus density radial parameter ($p$), the torus density angular parameter ($q$), the angle between the equatorial plane and the edge of the torus ($\Delta$), the ratio of the maximum to minimum radii of the torus ($R_{\rm max}/R_{\rm min}$), the viewing angle ($\theta$), and the AGN fraction of the total IR luminosity ($f_{\rm AGN}$), all of which are listed in Table \ref{Param}.
 
We modeled the X-ray emission using a combination of (i) an AGN with a power-law photon index ($\Gamma$) defined as $f_{\nu} \propto \nu^{1-\Gamma}$, (ii) low-mass X-ray binaries (LMXBs), (iii) high-mass X-ray binaries (HMXBs), and (iv) hot gas.
The X-ray emissions from the LMXBs and HMXBs were connected to the assumed SFH and SSP, and the hot-gas emission was modeled using free--free and free--bound emission from optically-thin plasma with $\Gamma$ = 1.
To connect the continuum emission from X-rays with other wavelengths, {\tt X-CIGALE} employs an empirical relation between $\alpha_{\rm OX}$ and $L_{\rm 2500\AA}$ \citep{Just}, where $L_{\rm 2500\AA}$ is the intrinsic (de-reddened) AGN luminosity density at 2,500 \AA, and $\alpha_{\rm OX}$ is the SED slope between 2,500 \AA\, and 2 keV.
The derived value of $\alpha_{\rm OX}$ is allowed to deviate from that expected from the $\alpha_{\rm OX}$--$L_{\rm 2500\AA}$  relation \citep{Just} by up to $\sim2\sigma$ (i.e., $|\Delta\,\alpha_{\rm OX}|_{\rm max} = |\alpha_{\rm OX} - \alpha_{\rm OX}\,(L_{\rm 2500\AA})|_{\rm max} = 0.2$), following \cite{Yang}.

Using the parameter settings listed in Table \ref{Param}, we fitted the stellar, nebular, AGN, SF, and X-ray components to at most 25 photometric points from X-ray to FIR wavelengths of the eFEDS-W4-X subsample.
Below, we explain the photometry used for each dataset employed for the SED fitting.
For the X-ray data, we used {\tt FluxCorr\_S} and {\tt FluxCorr\_H}, which are the absorption-corrected fluxes in the 0.5--2 and 2--10 keV bands, respectively \citep[see][]{Liu}.
For the UV data, we used {\tt FUV\_mag} and {\tt NUV\_mag}, which are the {\it GALEX} calibrated magnitudes in the FUV and NUV wavelengths\cite[see][]{Bianchi}.
For the optical (DECaLS) and MIR ({\it unWISE}) data in LS8, we used forced photometry at the positions of the optical counterparts \citep[see][]{Dey_19}. 
For the SDSS and HSC optical data, we used the {\tt cModel} magnitude, whose HSC five-band photometry was forced photometry \citep[see e.g.,][]{Bosch}.
For the optical--NIR data taken from KiDS, we used aperture-matched forced photometry at the positions of the optical counterparts ({\tt MAG\_GAAP\_u/g/r/i/z/y/j/h/ks}) \citep[see][]{Kuijken}.
For the NIR data taken from UKIDSS, we utilized the \cite{Petrosian} magnitudes ({\tt j/h/kPetroMag}) \citep[see also][]{Lawrence}.
For the FIR data, we used the default fluxes (i.e., {\tt FLUX\_65/90/140/160} for {\it AKARI} and {\tt F100/160/250/350/500B} for {\it Herschel}) \citep[see][]{Valiante,Yamamura}.
The flux densities at 250, 350, and 500 $\mu$m can be boosted, especially for faint sources (known as flux boosting or flux bias), which are caused by confusion noise and instrument noise. 
Hence, we corrected this effect using the correction term provided in Table 6 of \cite{Valiante}.
We corrected all the flux densities from X-ray to MIR wavelengths for galactic extinction.

Further, we examined the consistency of optical and NIR photometry for objects detected using the same band taken from different surveys (e.g., KiDS--VIKING $u$-band and SDSS $u$-band).
For the $u$-band magnitude, the median value of the difference between KiDS--VIKING and SDSS was 0.08 mag.
For the optical data, the median values of the differences between (i) KiDS--VIKING and LS8 ($g$, $r$, and $z$), (ii) HSC and LS8 ($g$, $r$, and $z$), and (iii) KiDS--VIKING and HSC ($g$, $r$, $i$, $z$, and $y$) were typically 0.09, 0.05, and 0.03 mag, respectively, which is acceptable in this work.
However, the median values of the differences between KiDS--VIKING and UKIDSS for the $J$, $H$, and $K$-bands were 0.12, 0.15, and 0.11 mag, respectively, which is a relatively large discrepancy.
Although the differences in optical and NIR photometry did not significantly affect the resulting physical values, as discussed in Sect.~\ref{R_SED}, we should consider the possible systematic differences in NIR photometry in this work.
 
Following \cite{Toba_19b}, we used the flux density at a given wavelength when SN was greater than 3 at that wavelength.
If an object was undetected, we placed 3$\sigma$ upper limits at those wavelengths\footnote{{\tt X-CIGALE} handles the SED fitting of photometric data with upper limits by employing the method presented by \cite{Sawicki}. This method computes $\chi^{2}$ by introducing the error function (see Equations [15] and [16] in \cite{Boquien}.}.
Although the photometry employed in each catalog is different, they may trace the total flux densities.
Therefore, the influence of differences in the photometry is expected to be small.
Nevertheless, it is worth addressing whether physical properties are actually estimated in a reliable way given the uncertainties in each source of photometry, which we discuss in Sect.~\ref{R_SED}.

\subsection{2D image decomposition}
\label{s_2D}

We performed 2D image analysis in the same manner as in \cite{Li}, which involved an automated fitting routine based on the image modeling tool {\tt Lenstronomy} \citep{Birrer,Birrer_18} to decompose the image into a point source and a host-galaxy component.
A full description is provided in \cite{Ding} and \cite{Li}.
We briefly summarize the analysis here.
For image decomposition, we used a cut-out image obtained from the HSC $i$-band\footnote{Because the $i$-band data provide the highest image quality among the HSC five-band data for the typical seeing of 0$\farcs$6 \citep{Aihara18a}, the $i$-band is ideal for 2D image analysis.}, for which the image size was automatically optimized in the procedure.

We performed 2D brightness profile fitting for the background-subtracted cut-out images using a point spread function (PSF) (for the point source) and a single elliptical S\'ersic model (for the host galaxy) to determine galactic structural properties, such as the S\'ersic index ($n$) and half-light radius ($R_{\rm e}$) as well as the flux densities of the central point source and the host galaxy (see Equation [1] in \citealt{Li}). 
The best-fit model image was obtained using $\chi^2$ minimization based on the particle swarm optimization algorithm \citep{Kennedy}.
The 1$\sigma$ error of the best-fit structural parameters and the flux density of each component could be inferred through the Markov-chain Monte Carlo (MCMC) sampling procedure in {\tt Lenstronomy} \citep{Foreman}.
However, we eventually employed a 10\% error for each quantity because the derived errors based on the MCMC procedure were found to be unrealistically small, possibly due to adopting a fixed PSF in our 2D profile fitting \citep[see Sect.~3.1 in][]{Li}.

Fig.~\ref{fig_decomp} presents examples of quasar-host decomposition based on the HSC $i$-band image.
We found that the reduced $\chi^2$ ($\chi_{\rm decomp}^2$) of the profile fitting worsened for objects at $z < 0.1$, probably because the radial profile of a low-$z$ object is often difficult to explain with a single S\'ersic model \citep{Li}.
Moreover, it was difficult, even when using the HSC data, to decompose the host emission for objects at $z > 1$ given the angular resolution of the HSC.
Following \cite{Li}, we used products (e.g., host flux densities) for objects at $0.2 < z < 0.8$ with reduced $\chi_{\rm decomp}^2 < 5$ to ensure reliable host properties (see Sect.~\ref{D_decomp}).

\subsection{X-ray stacking analysis} 
\label{s_st}

Among the 7,088 objects in the eFEDS--W4-nonX sample, 5,822 objects were in the survey footprint of the HSC-SSP.
Because 4,288/5,822 ($\sim$74\%) sources did not have $z_{\rm spec}$, we computed photometric redshifts ($z_{\rm photo}$) for them.
We utilized $z_{\rm photo}$ calculated using the machine learning code Deep Neural Network Photo-z ({\tt DNNz}; Nishizawa et al., in prep.).
The \texttt{DNNz} comprises multilayered perceptrons with four fully connected hidden layers, with each layer having 100 nodes. For the inputs, we have used the cmodel fluxes and PSF-convolved aperture fluxes for five HSC broadband filters, translated into \texttt{asinh} magnitudes \citep{Lupton}, as well as the sizes derived from second-order moments for each filter. 
The output is the probability of an individual galaxy being in a certain redshift bin, with redshifts divided linearly from $z=0$--$7$. With the spectroscopic-redshift sample and the high-quality photometric-redshift sample in COSMOS, which are summarized in \cite{Nishizawa}, the code is trained so that the output probability approaches a Dirac delta function at the true redshift. 

\begin{figure}
\centering
\includegraphics[width=0.4\textwidth]{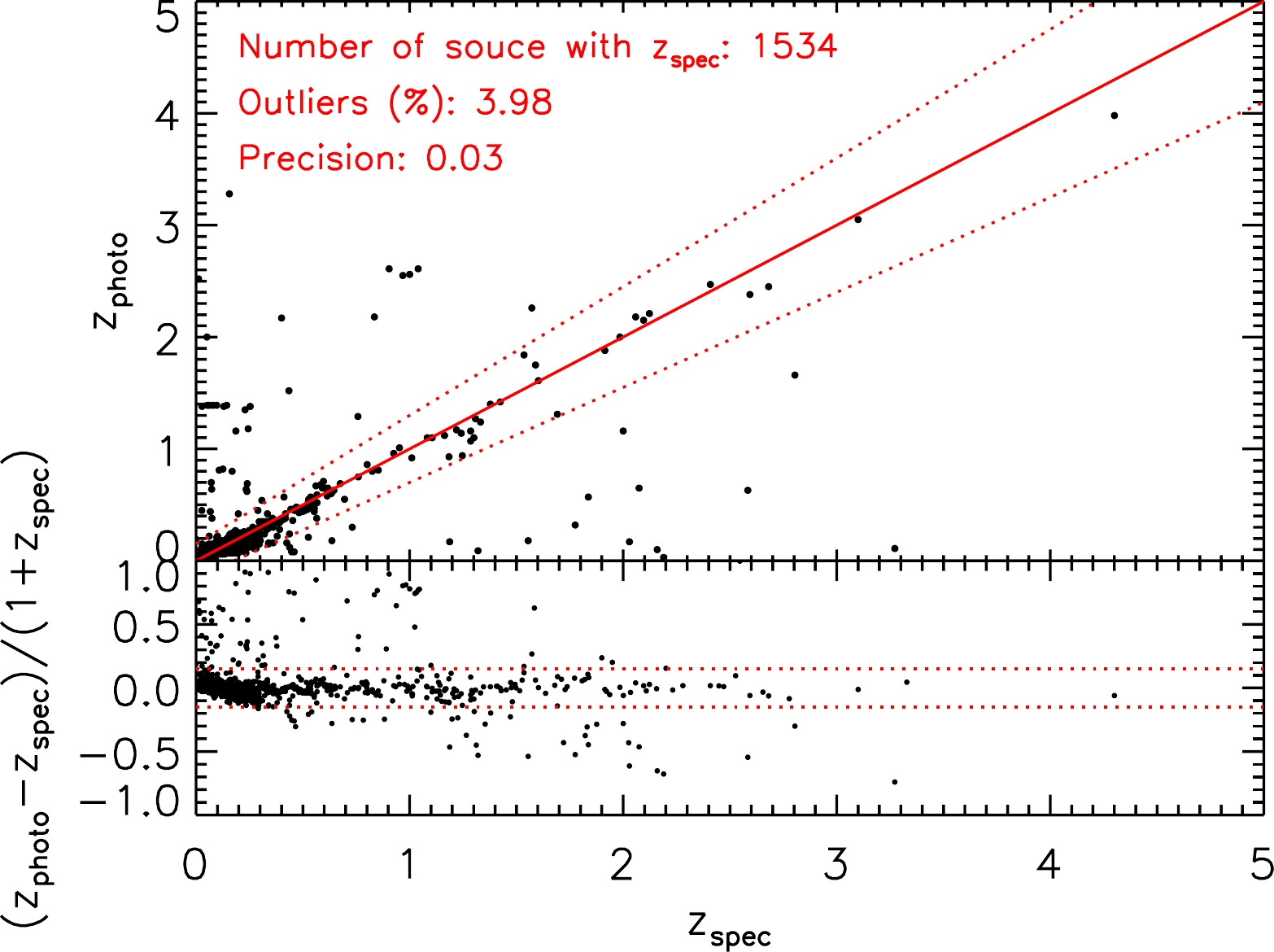}
\caption{Comparison between $z_{\rm spec}$ and $z_{\rm photo}$ for the eFEDS--W4-nonX sample. The red solid line represents $z_{\rm spec} = z_{\rm photo}$. The red dotted lines denote $\Delta z$/(1+$z_{\rm spec}$) = $\pm 0.15$. The bottom panel shows $\Delta z$/(1+$z_{\rm spec}$) as a function of $z_{\rm spec}$.}
\label{fig_photoz}
\end{figure}

Fig.~\ref{fig_photoz} shows a comparison between $z_{\rm spec}$ and $z_{\rm photo}$ for the eFEDS--W4-nonX sample, in which we compiled $z_{\rm spec}$ from the SDSS DR15, GAMA DR3, and WiggleZ DR1.
The resulting accuracy of $z_{\rm photo}$, the normalized median absolute deviation defined as $\sigma_{\Delta z/(1+z_{\rm spec})}$  = 1.48 $\times$ median($|\Delta z|$/(1+$z_{\rm spec}$)) \citep[e.g.,][]{Ilbert}, is 0.03.
The outlier fraction (i.e., the fraction of sources with $|\Delta z|$/(1+$z_{\rm spec}) > 0.15$) is 3.98\%, which is acceptable in this work (see also Sect.~\ref{Xana}).
However, we have verified the performance of $z_{\rm photo}$ only for the objects at $z < 4$.
Therefore, we only considered the objects at $z < 4$ to ensure the reliability of $z_{\rm photo}$ for these high-$z$ sources (see Fig.~\ref{fig_zhist}).
Moreover, we adopted {\tt brightblob}\footnote{\url{https://www.legacysurvey.org/dr8/bitmasks/#id1}}$\neq$ 3 to extract extragalactic objects with reliable redshifts (see also Sect.~\ref{s_photoz}). 
Accordingly, 4,441 objects (hereafter referred to as the eFEDS--W4-nonX subsample) remained for the stacking analysis.
The sample was separated into five subsamples using the following redshift bins: (i) $z < 0.1$, (ii) $0.1 < z < 0.25$, (iii) $0.25 < z < 0.5$, (iv) $0.5 < z < 1.0$, and (v) $z > 1.0$, and we then stacked each subsample as follows.

We first masked the previously detected sources using the position from the catalog \citep{Salvato_21} in the X-ray images (30$\arcsec$ radius for point source and a spatial extent for extended sources). 
This ensured that we could properly characterize the background. We performed masking on four bands, i.e., full (0.2--10 keV), soft (0.2--0.6 keV), mid (0.6--2.3 keV), and hard (2.3--5.0 keV) \citep[see][]{Brunner}.
We then stacked images in all four bands by taking the mean of the inverse exposure weighted $2\arcmin \times 2\arcmin$ cutouts centered on each {\it WISE} position, essentially creating count-rate stacked images. 
To not overly weigh sources at the edges of the field, we required a minimum exposure time of 180 s, which reduced the number of stacked sources by 3\%--5\%, with the larger percentage in the hard band, which experienced a lot of vignetting (see Sect.~\ref{R_stack}).

Thereafter, we attempted to detect the source in the stacked image by PSF-matching the center to the background. 
We selected a boundary SN of 1.8 as our boundary, which corresponded to a matched detection significance of 5.0. 
Lastly, we determined the (limiting) fluxes above the background using the following energy conversion factors: full, $5.03 \times 10^{11}$; soft, $9.61\times 10^{11}$; mid, $1.06 \times 10^{12}$; and hard, $1.13\times10^{13}$. 
They were calculated using an absorbed power-law model with $\log N_{\rm H}$ = 20 cm$^{-2}$ and $\Gamma$ of 1.7.

\section{Results}
\label{R}

As described in Sects.~\ref{s_photoz} and \ref{s_st}, we conservatively narrowed down the eFEDS--W4 sample to 360 objects in the eFEDS--W4-X subsample and to 4,441 objects in the eFEDS--W4-nonX subsample with $z < 4$ to ensure the reliability of the determined physical quantities (see also Fig.~\ref{fig_SS}).

\subsection{Results of X-ray spectral analysis}
\label{R_X}

\begin{figure}
\centering
\includegraphics[width=0.35\textwidth]{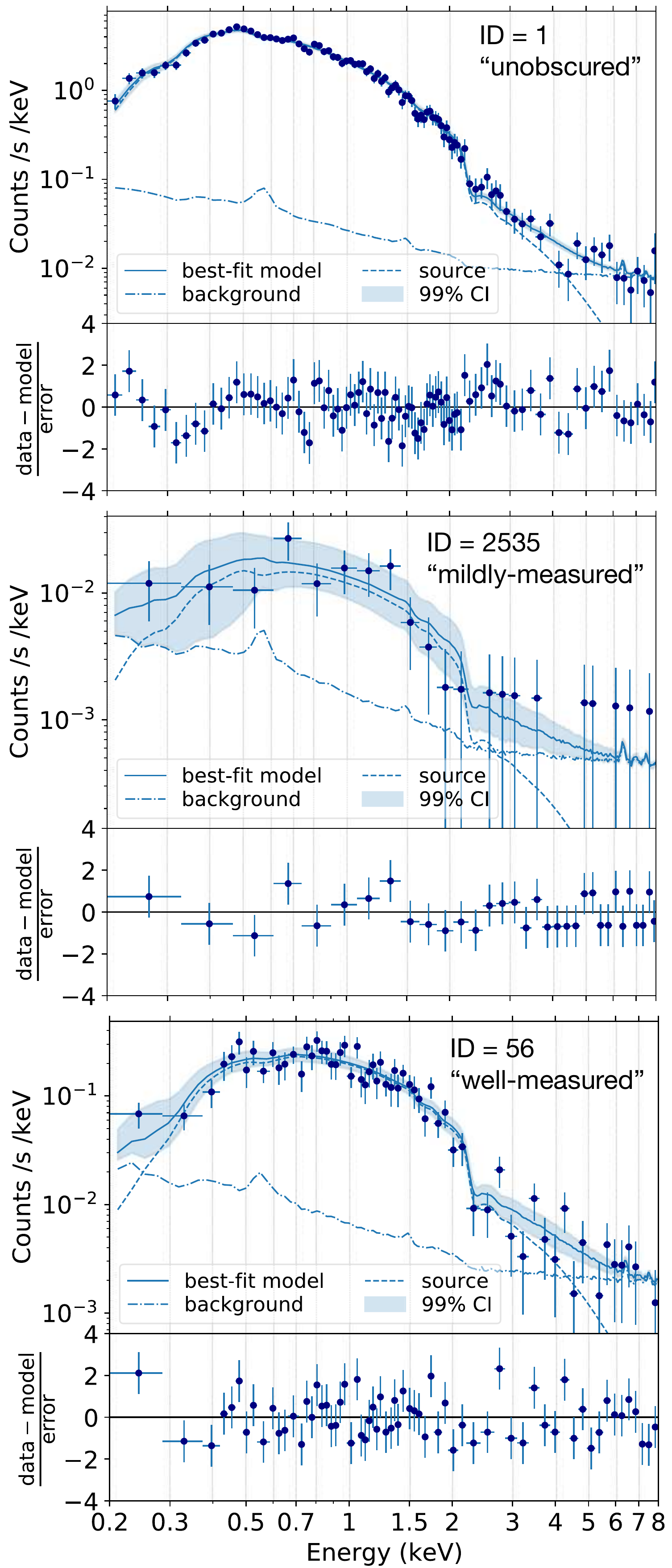}
\caption{Examples of the X-ray spectra from {\it eROSITA}. The best-fit model and its 99\% confidence interval are displayed as blue-shaded regions. For each source, the lower panel shows comparison of the data with the best-fit model in terms of (data$-$model)/error, where error is calculated as the square root of the model predicted number of counts.}
\label{fig_X}
\end{figure}
\begin{figure}
\centering
\includegraphics[width=0.4\textwidth]{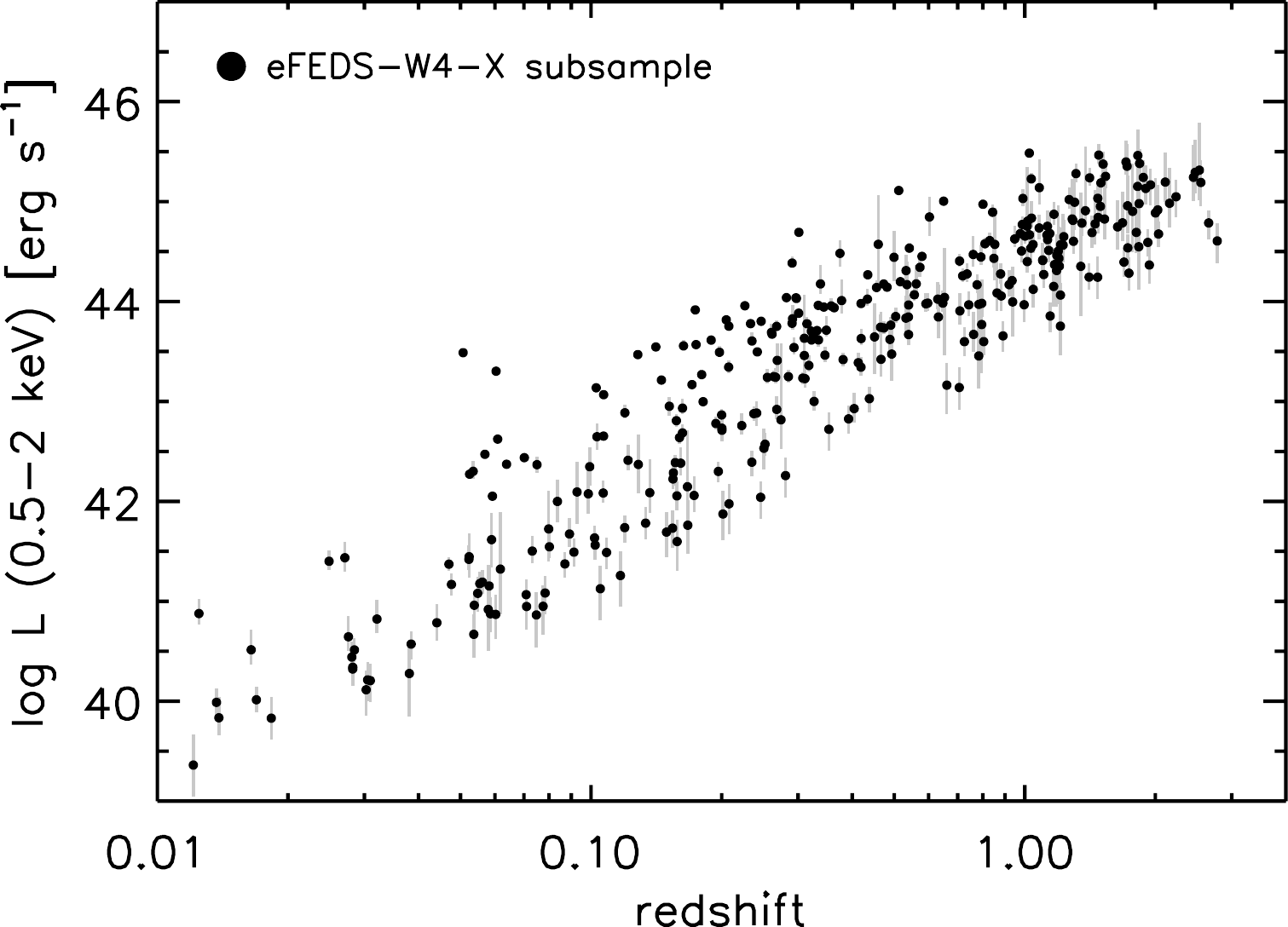}
\caption{Absorption-corrected X-ray luminosity in the 0.5--2 keV band for the 337 objects in the  eFEDS--W4-X subsample as a function of redshift.}
\label{fig_z_Lx_s}
\end{figure}

\begin{figure}
\centering
\includegraphics[width=0.4\textwidth]{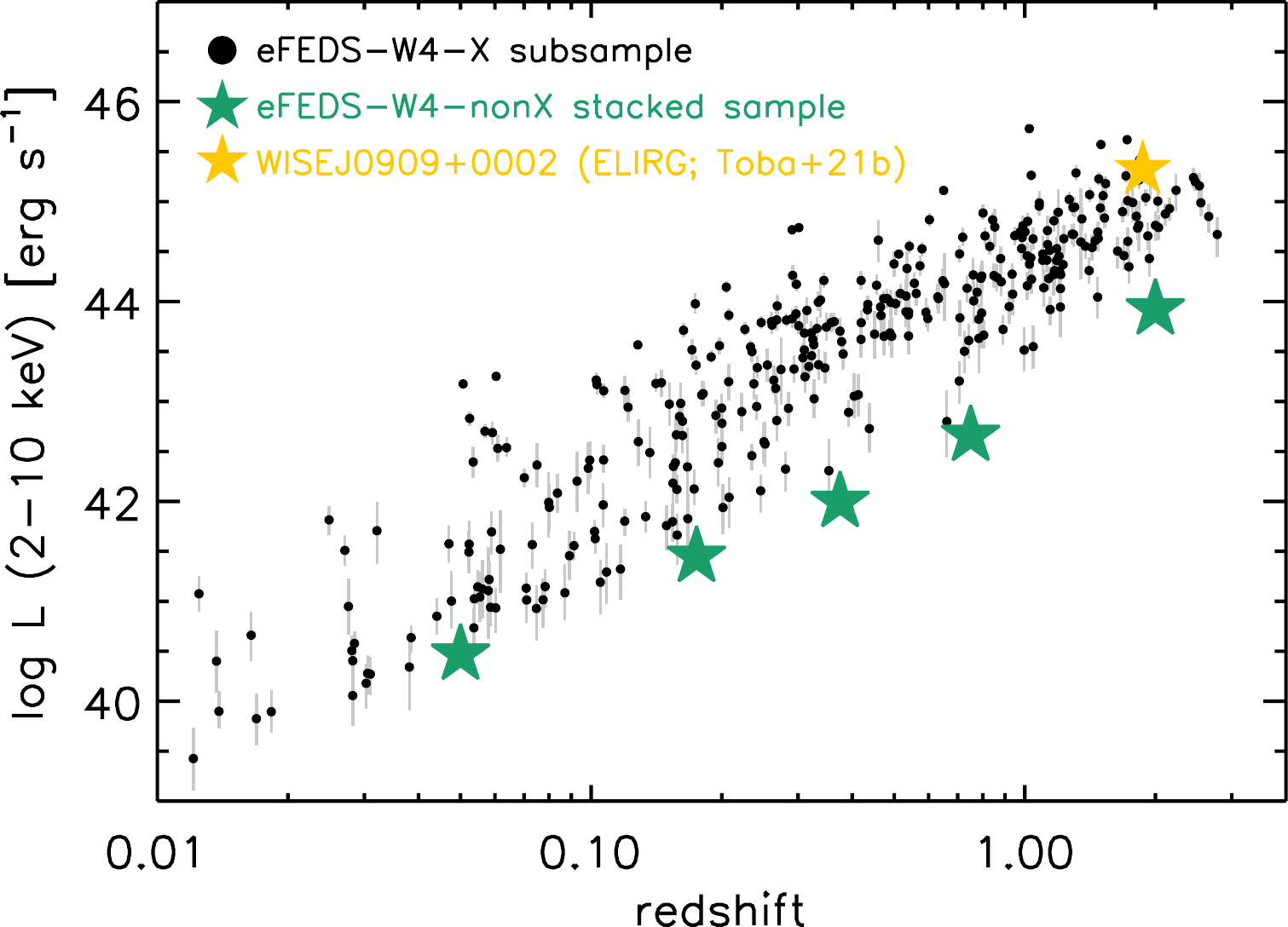}
\caption{Absorption-corrected X-ray luminosity in the 2--10 keV band for the 339 objects in the eFEDS--W4-X subsample as a function of redshift. The green stars represent the eFEDS--W4-nonX subsample used in the X-ray stacking analysis, for which we calculated $L_{\rm X}$ by assuming $N_{\rm H}$ = 0 cm$^{-2}$ and $\Gamma = 2.0$. The yellow star denotes the ELIRG (WISEJ0909+0002) reported in \cite{Toba_21b}.}
\label{fig_z_Lx_h}
\end{figure}
\begin{figure}
\centering
\includegraphics[width=0.4\textwidth]{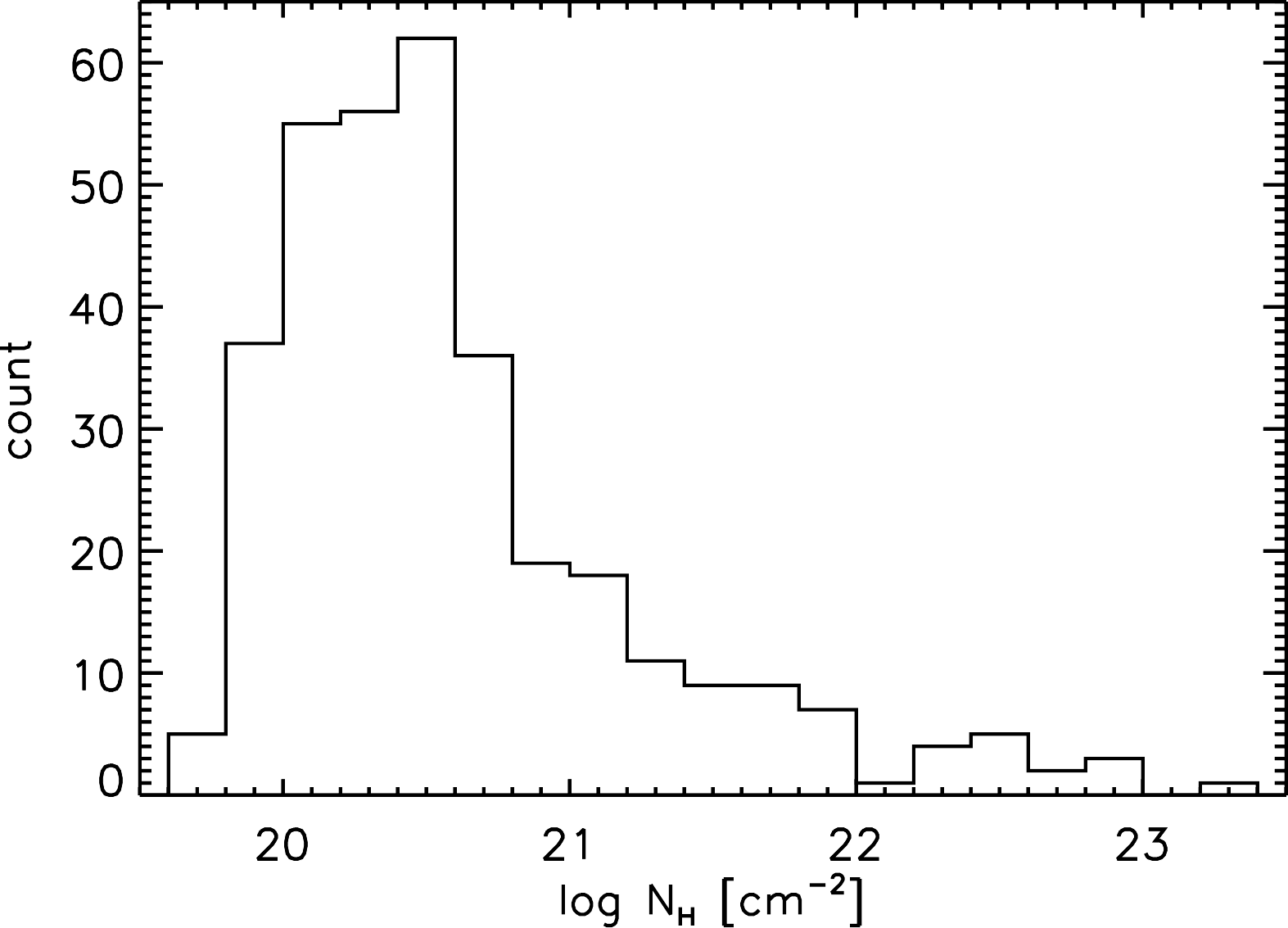}
\caption{Histogram of hydrogen column density ($N_{\rm H}$) measured from the {\it eROSITA} X-ray spectra for 339 objects.}
\label{fig_NH}
\end{figure}
\begin{figure*}
\centering
\includegraphics[width=0.85\textwidth]{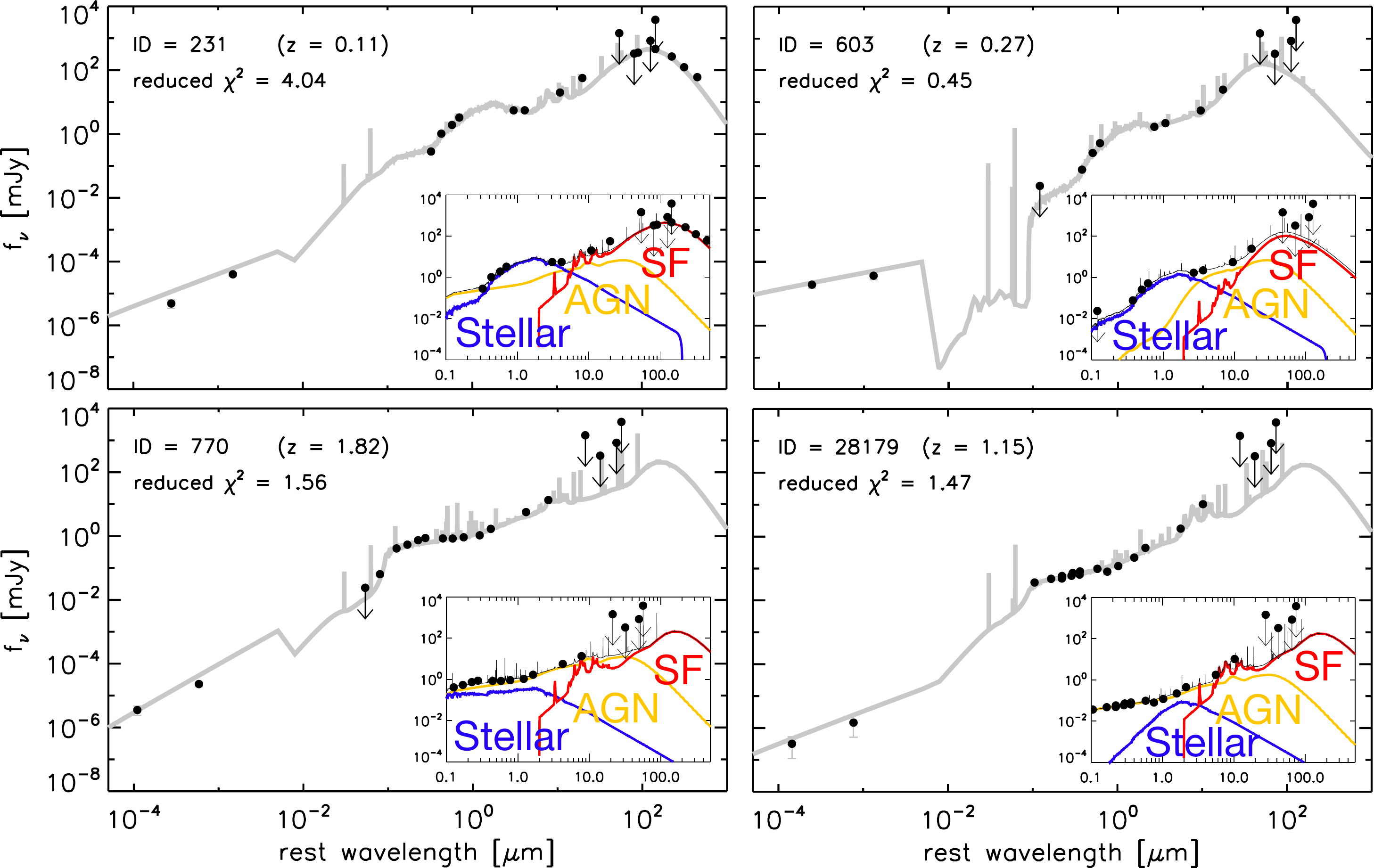}
\caption{Examples of SED fitting. The black points are photometric data, and the gray solid line represents the resulting best-fit SED. The inset shows the SED at 0.1--500 $\mu$m, with the contributions from the stellar, AGN, and SF components to the total SED shown as blue, yellow, and red lines, respectively.}
\label{fig_SED}
\end{figure*}

Fig.~\ref{fig_X} presents examples of the X-ray spectra.
We estimated the absorption-corrected X-ray luminosities in the rest-frame soft (0.5--2 keV) band ($L_{\rm X}$ (0.5--2 keV)) and hard (2--10 keV) band ($L_{\rm X}$ (2--10 keV)) for all the 360 objects.
However, for 21 objects (with a class of {\tt uninformative}), $L_{\rm X}$ has large uncertainties because the shapes of the X-ray spectra have been fixed \citep[see][]{Liu}. We therefore exclude them from the following analysis.
The luminosities of these $360 - 21 = 339$ objects are presented as a function of redshift in Figs.~\ref{fig_z_Lx_s} and \ref{fig_z_Lx_h}.
The mean values and standard deviations of these luminosities are $\log\,L_{\rm X}$ (0.5--2 keV) = $43.4 \pm 1.41$ erg s$^{-1}$ and $\log\,L_{\rm X}$ (2--10 keV) = $43.4 \pm 1.37$ erg s$^{-1}$.
If we consider objects with $\log\,L_{\rm X}$ (2--10 keV) $>$ 42 erg s$^{-2}$ as AGN \citep[e.g.,][]{Barger}, 278/339 ($\sim$82\%) objects satisfy the criterion.
We find that approximately  86\% and 14\% of X-ray AGN are classified as X-ray type 1 and type 2 AGN according to the best-fit model (see Sect.~\ref{Xana}).

We present $N_{\rm H}$ for 339 objects in the eFEDS--W4-X subsample in Fig.~\ref{fig_NH}.
The mean value and standard deviation of $\log\,N_{\rm H}$ for our AGN sample is $\log\,N_{\rm H} = 20.6 \pm 0.69$ cm$^{-2}$, and the mean value and standard deviation of the photon index is $\Gamma = 2.0 \pm 0.3$; both are in good agreement with a previous work on X-ray AGN \citep{Trump}.

\subsection{Results of SED fitting}
\label{R_SED}

Fig.~\ref{fig_SED} shows examples of the SED fits obtained with {\tt X-CIGALE}.
We confirmed that 197/360 ($\sim$55\%) objects reduced $\chi^2 \leq 3$, while 263/360 ($\sim$73\%) objects have reduced $\chi^2 \leq 5$.
This suggests that the data are moderately well fitted by {\tt X-CIGALE} with a combination of stellar, nebular, AGN, SF, and X-ray components.
Hereafter, we focus on the subsample of 263 eFEDS--W4-X sources with reduced $\chi^2$ smaller than 5 for the SED fitting, in the same manner as in \cite{Toba_19b,Toba_20c}.

\begin{figure}
\centering
\includegraphics[width=0.4\textwidth]{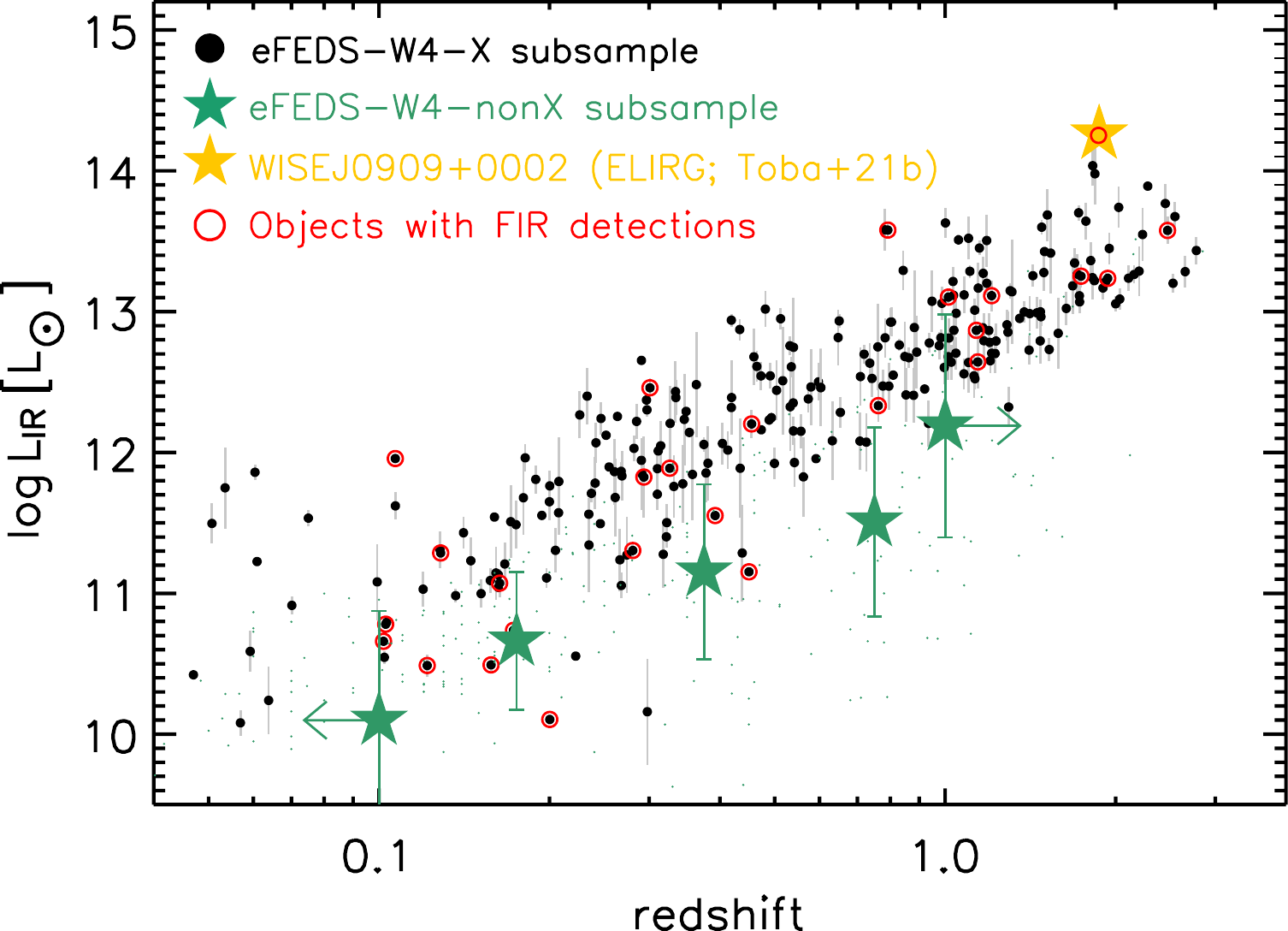}
\caption{IR luminosity as a function of redshift. The green stars represent the eFEDS--W4-nonX subsample, for which the values of $L_{\rm IR}$ are the mean and standard deviation for sources (indicated by green dots) in each redshift bin, as given in Table \ref{T_stack}. The yellow star represent the ELIRG (WISEJ0909+0002) reported in \cite{Toba_21b}. Objects with red circles have FIR detections from {\it Herschel}/{\it AKARI}.} 
\label{fig_z_LIR}
\end{figure}

Fig.~\ref{fig_z_LIR} shows the IR luminosity ($L_{\rm IR}$).
$L_{\rm IR}$ is traditionally defined as the luminosity integrated over the wavelength range 8--1000 $\mu$m \citep[e.g.,][]{Sanders,Chary}. However, because this definition includes contributions from stellar emissions, we did not adopt any wavelength boundary for the integration range. Instead, we employed a physically oriented approach to estimate $L_{\rm IR}$ as a function of redshift based on {\tt X-CIGALE} by considering the energy re-emitted by dust that has absorbed UV/optical emission from SF/AGN \citep[see e.g.,][]{Toba_20b,Toba_21b}.
We found that approximately 67\% of our samples are ultraluminous IR galaxies \citep[ULIRGs:][]{Sanders} and 24\% are hyperluminous IR galaxies \citep[HyLIRGs:][]{Rowan-Robinson}, with $L_{\rm IR} > 10^{12}$ and $>10^{13}$ $L_{\sun}$, respectively.
This suggests that a large fraction of X-ray-detected {\it WISE} 22 $\mu$m sources are ULIRGs and may even be HLIRGs.
Furthermore, we discovered two candidates (0.8\% of our sample) that are extremely luminous IR galaxies \citep[ELIRGs:][]{Tsai}, with $L_{\rm IR}$ greater than $10^{14}$ $L_{\sun}$; indeed, \cite{Toba_21b} confirm that the object named WISEJ0909+0002 at $z_{\rm spec}$ = 1.871 is an ELIRG.
However, we note that {\tt CIGALE} would overestimate FIR emission from SF for objects that do not have FIR photometry \citep[e.g.,][]{Masoura}.
Only 26/263 objects in the eFEDS--W4--X subsample were detected by {\it Herschel} and/or {\it AKARI} (see objects with red circles in Fig.~\ref{fig_z_LIR}).
The remaining 237 sources had upper limits given by {\it Herschel} and {\it AKARI}, which would be insufficient to constrain the FIR SEDs.
We confirmed that the mean value of $L_{\rm IR}$ for these objects without FIR detections is $\log\,(L_{\rm IR}/L_{\sun}) = 12.4$, which is approximately 0.4 dex larger than that of the objects with FIR detections, which should be kept in mind as the potential systematic uncertainty of $L_{\rm IR}$ (and SFR) in this work.

\begin{figure}
\centering
\includegraphics[width=0.45\textwidth]{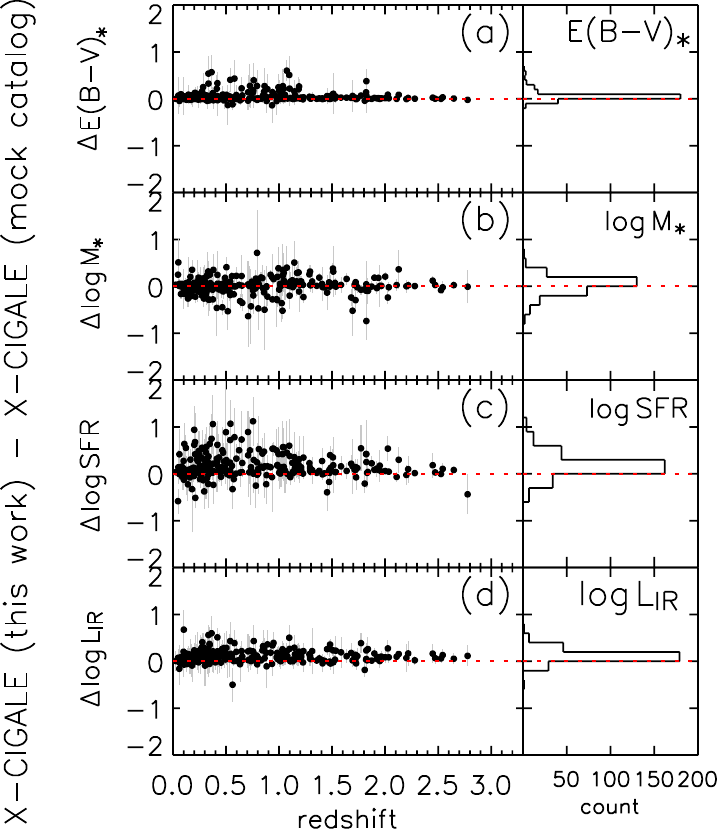}
\caption{Differences in $E(B-V)_{*}$, the stellar mass ($M_*$), the SFR, and $L_{\rm IR}$ derived from {\tt X-CIGALE} in this work and those derived from the mock catalog. (a) $\Delta$$E(B-V)_{*}$, (b) $\Delta$$\log\, M_{*}$, (c) $\Delta$$\log$ SFR, and (d) $\Delta$$\log \,L_{\rm IR}$ as functions of redshift. The right-hand panels display histograms of each quantity. The red dotted lines correspond to $\Delta$ = 0.}
\label{fig_mock}
\end{figure}

We performed a mock analysis to check whether the physical properties can be reliably estimated in the same manner as in \cite{Toba_19b,Toba_20c}.
To create the mock catalog, {\tt X-CIGALE} first uses the photometric data for each object based on the best-fit SED and then modifies the photometry for each one by adding a value taken from a Gaussian distribution with the same standard deviation as the observation. 
This mock catalog is then analyzed in exactly the same way as for the original observations (i.e., the mock analysis uses the same simulation model as the fitting model) \citep[see][for more details]{Boquien}.
This analysis also enables us to examine the influence of photometric uncertainties on the derived physical quantities (see Sect.~\ref{s_SED}).

Fig.~\ref{fig_mock} presents the differences in $E(B-V)_{*}$, the stellar mass, the SFR, and $L_{\rm IR}$ derived from {\tt X-CIGALE} in this work and those derived from the mock catalog as a function of redshift.
The mean values are $\Delta\, E(B-V)_{*}=0.05$, $\Delta\, \log\, M_{*}=0.004$, $\Delta\, \log$ SFR $=0.16$, and $\Delta\, \log \,L_{\rm IR}=0.11$.
Although there is no significant dependence on redshift, the relatively large offsets for SFR and $L_{\rm IR}$ suggest that these quantities may be sensitive to photometric uncertainties. 
This may be a limitation of our SED fitting method given the limited number of data points.
This possible uncertainty should be kept in mind in the following discussion.

\subsection{Results of X-ray stacking analysis}
\label{R_stack}

\begin{table*}
\caption{X-ray fluxes of the stacked eFEDS--W4-nonX subsample.} 
\label{T_stack}
\centering
\begin{tabular}{lcccccc}
\hline \hline
\multirow{2}{*}{redshift range} 									& 
\multirow{2}{*}{N}													&	
\multicolumn{4}{c}{$f_{\rm X}$ ($10^{-16}$ erg s$^{-1}$ cm$^{-2}$)} &	\\
\cline{3-6}
				&	&	0.2--0.5 keV	&	0.6--2.3 keV	&	2.3--5.0 keV	&	 0.2--10 keV  	\\
\hline
$z$ < 0.1			&	950		&	$1.3 \pm 0.6$	& 
									$3.6 \pm 0.7$ 	& 
									$<$ 18.8\tablefootmark{a}	 		&	
									$5.8 \pm1.5$ 	\\
0.1 < $z$ < 0.25	&	1325	&	$0.9 \pm 0.5$	& 
									$2.4 \pm 0.7$ 	& 
									$<$ 16.7\tablefootmark{a}	 		&	
									$3.8 \pm 1.2$	\\	
0.25 < $z$ < 0.5	&	774		&	$0.9 \pm 0.6$	& 
									$1.5 \pm 0.8$ 	& 
									$<$ 22.2\tablefootmark{a}	 		&	
									$2.9 \pm 1.5$	\\
0.5 < $z$ < 1.0		&	876		&	$0.8 \pm 0.6$	& 
									$1.3 \pm 0.8$ 	& 
									$<$ 21.9\tablefootmark{a}	 		&	
									$3.1 \pm 1.5$	\\																 $z$ > 1.0			&	513		&	$<$ 2.3\tablefootmark{a}				& 
									$<$ 2.9\tablefootmark{a}	 		& 
									$<$ 28.8\tablefootmark{a}	 		&	
									$<$ 5.7\tablefootmark{a}			\\							
\hline					
\end{tabular}
\tablefoot{
\tablefoottext{a}{3$\sigma$ upper limit.}
}
\end{table*}

\begin{figure*}
\centering
\includegraphics[width=0.8\textwidth]{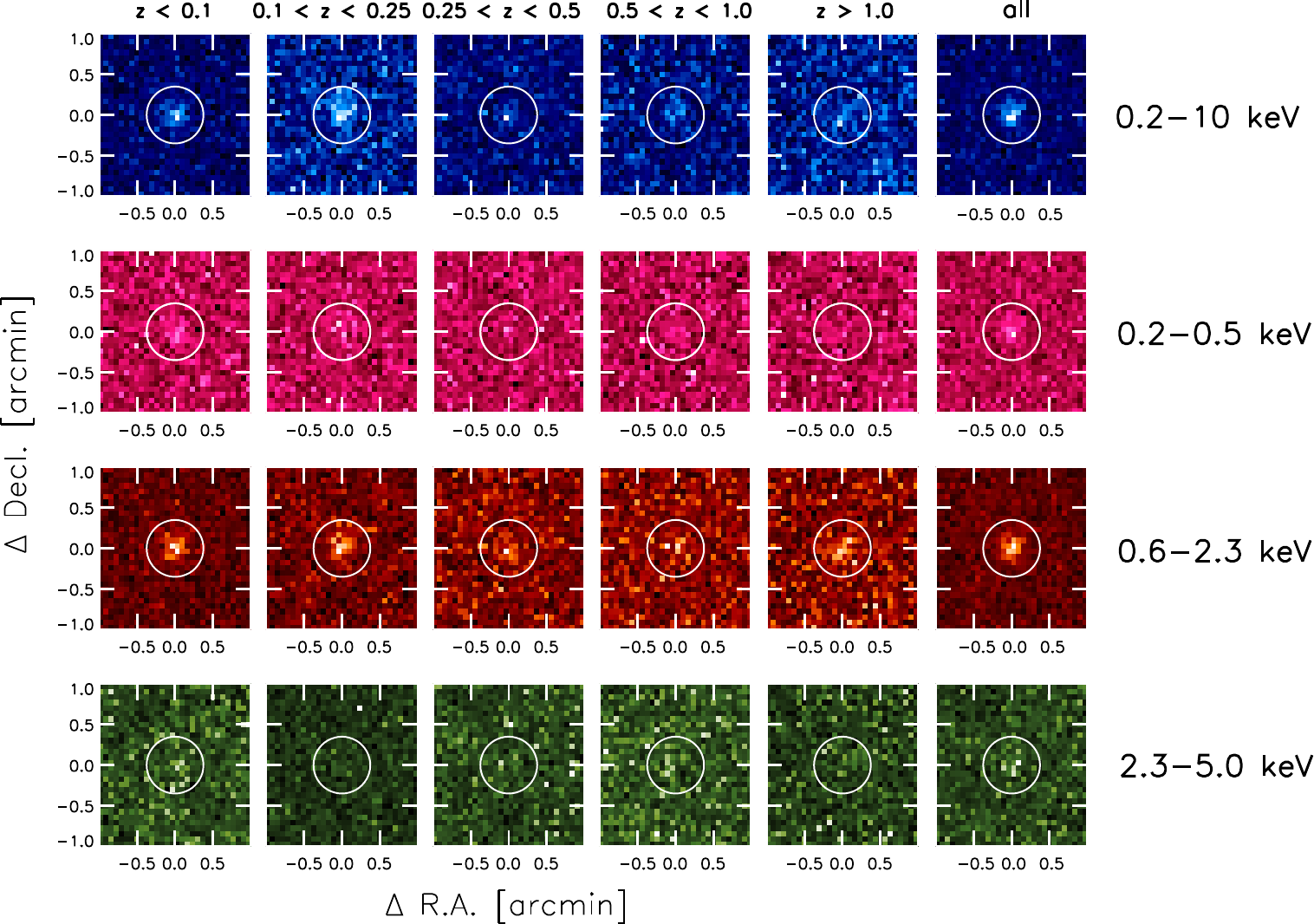}
\caption{Stacked X-ray images ($2\arcmin \times 2\arcmin$) of the eFEDS--W4-nonX samples as a function of redshift bin. From left to right, the bins are $z < 0.1$, $0.1 < z < 0.25$, $0.25 < z < 0.5$, $0.5 < z < 1.0$, $z > 1.0$, and all the samples. The energy bands (in the observed frame) of the subsample are 0.2--10, 0.2--0.5, 0.6--2.3, and 2.3--5.0 keV from top to bottom.}
\label{fig_stack}
\end{figure*}

Figure \ref{fig_stack} shows the stacked images in the observed frame of all the eFEDS--W4-nonX samples as functions of redshift.
The resulting X-ray flux in each band is tabulated in Table~\ref{T_stack}.
X-ray photons are detected in the 0.2--0.5-, 0.6--2.3-, and 0.2--10-keV bands for objects at $z < 1$, while 3$\sigma$ upper limits are obtained for the 2.3--5.0-keV band regardless of redshift.
Using the stacked spectra at these five redshift bins, we estimated the mean intrinsic (absorption-corrected) 2--10 keV luminosities of the eFEDS--W4-nonX subsamples. 
To convert the observed count rates into luminosities, we consider an absorbed power-law model with a 1\% unabsorbed scattered component, assuming $\Gamma = 2.0$ and four different values of absorption at the source
redshift, $N_{\rm H}=0$, $10^{22}$, $10^{23}$, and $10^{24}$ cm$^{-2}$, in addition to the Galactic absorption of $N_{\rm H} = 2.58\times10^{20}$ cm$^{-2}$ \citep{Liu}. 
For simplicity, we ignore other components such as optically-thin thermal emission and emissions from X-ray binaries in the host galaxies. 
Thus, our estimates of the AGN luminosities must be taken to be upper limits. 
We basically refer to the count rates in the 0.6--2.3 keV band, where {\it eROSITA} is the most sensitive and all the stacked spectra are significantly detected. 
If the 3$\sigma$ upper limit of the 2.3--5.0 keV count rate gives a smaller luminosity than the estimate from the 0.6--2.3 keV count rate, then we adopt the former value as an upper limit (which is the case only for $z<0.1$ and $N_{\rm H} = 10^{23}$ cm$^{-2}$). 
The results are summarized in Table \ref{T_stack_Lx}. 
It is noteworthy that the presence of obscured AGN with $L_{2-10} > 10^{42}$ erg s$^{-1}$ in these X-ray-undetected
sources is possible, although the inferred mean AGN luminosity is highly dependent on the assumed absorption column density.

We present the resulting values of $L_{\rm X}$ for the eFEDS--W4-nonX stacked sample as a function of redshift in Fig.~\ref{fig_z_Lx_h}, where $L_{\rm X}$ is plotted assuming $N_{\rm H}$ = 0 cm$^{-2}$.
We observe that the values of $L_{\rm X}$ for this stacked sample are smaller than those for the eFEDS--W4-X sample.
Given the fact that the majority of the eFEDS--W4-X sample is expected to be unobscured AGN (see Sect.~\ref{R_X} and Fig.~\ref{fig_NH}), this result indicates that the estimates of $L_{\rm X}$ for the eFEDS--W4-nonX stacked sample are reasonable.
Fig.~\ref{fig_z_LIR} presents the resulting values of $L_{\rm IR}$ for the eFEDS--W4-nonX subsample as a function of redshift.
Because $L_{\rm IR}$ can be estimated for individual sources in the sample (see below), we plotted the mean values and standard deviations of $L_{\rm IR}$ for each redshift bin in addition to plotting individual sources.
We can see that the values of $L_{\rm IR}$ for the eFEDS--W4-nonX subsample are smaller than those for the eFEDS--W4-X subsample, although there is large dispersion.
This is possibly because our eFEDS--W4-X subsample was limited to objects with $z_{\rm spec}$; thus, their $L_{\rm IR}$ would be biased toward higher luminosity, which is confirmed via the stacked SED described below.

\begin{figure}
\centering
\includegraphics[width=0.45\textwidth]{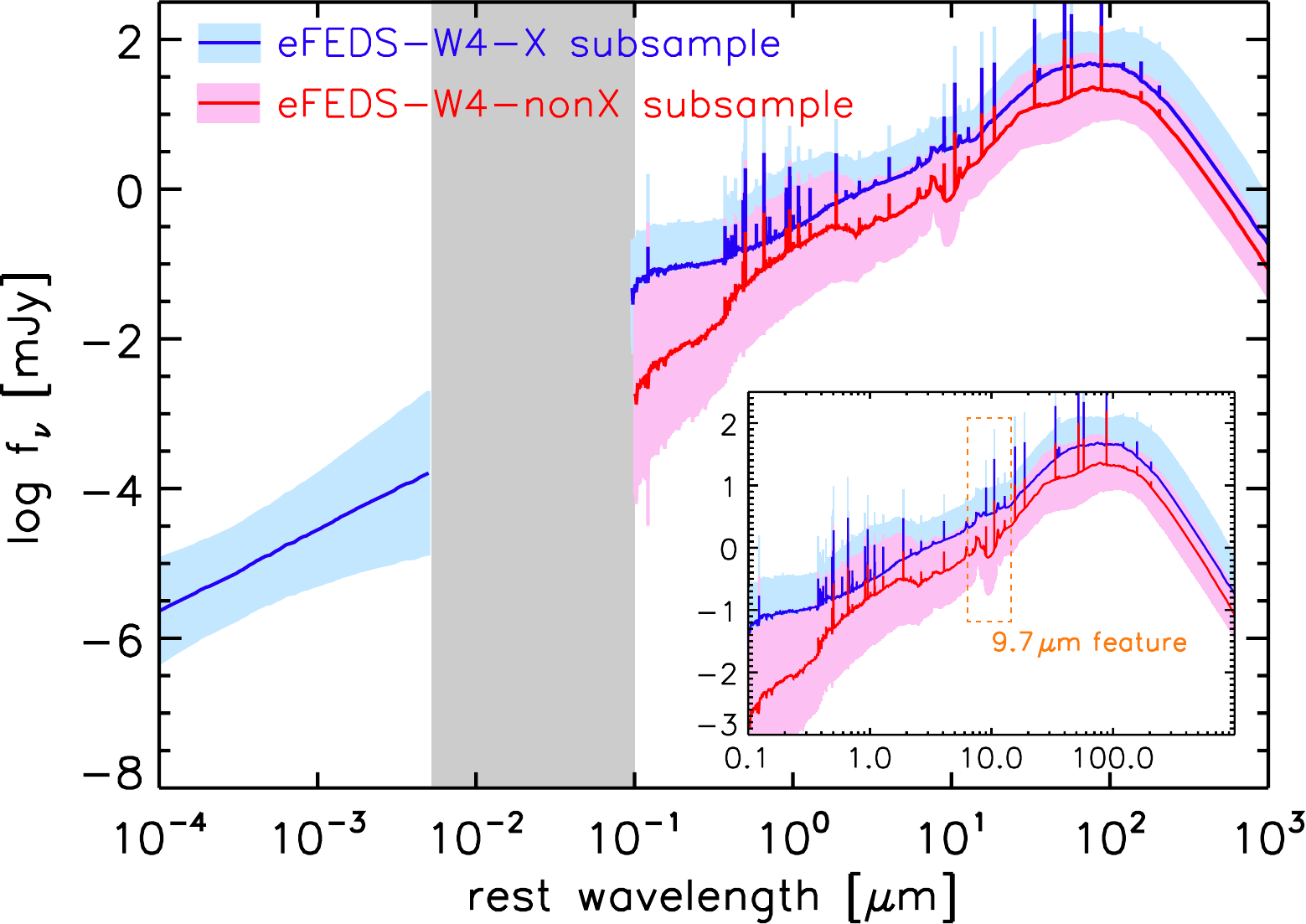}
\caption{Composite SEDs of the eFEDS-W4-X subsample (blue) and eFEDS-W4-nonX subsample (red). The shaded regions represent the standard deviations of the stacked SED medians. There are no data points in the gray region, and composite SEDs are excluded. The inset displays a magnified view of the rest-frame optical-to-FIR SEDs.}
\label{fig_comp_SED}
\end{figure}

To compare the shapes of the SEDs between the eFEDS--W4 samples with and without {\it eROSITA} detections, we performed SED fitting for the eFEDS--W4-nonX subsample.
We compiled the multiwavelength data and performed SED fitting to the UV--FIR data in the same manner as described in Sects.~\ref{s_multi} and \ref{s_SED}.
We created the composite SED by stacking the best-fit SEDs for all the objects.
The composite SEDs of the eFEDS--W4-X and eFEDS--W4-nonX subsamples are displayed in Fig.~\ref{fig_comp_SED}, where we do not use $\sim$0.005--0.1 $\mu$m for stacking the SEDs because there are no data points in that wavelength range.
No significant differences are observed in the NIR and FIR SEDs among these sources.

In the optical region, however, the eFEDS--W4-X subsample exhibits a power-law SED caused by emission from the AGN accretion disk, while the eFEDS--W4-nonX subsample exhibits Balmer and 4000 \AA\, break features caused by emission from the AGN host.
In addition, the strength of the silicate feature at 9.7 $\mu$m (i.e., $\tau_{\rm 9.7}$) is larger for the eFEDS--W4-nonX subsample than for the eFEDS--W4-X subsample (see the inset of Fig.~\ref{fig_comp_SED}).
The power-law feature of the eFEDS--W4-X subsample is consistent with the fact that the majority of the sample comprises X-ray type 1 AGN (see Sect. \ref{R_X}).
However, some eFEDS--W4-nonX sources may be dust-obscured type 2 AGN, including CT-AGN (see Sect.~\ref{D_NH_EDD}). 
This is supported by the relatively strong 9.7 $\mu$m silicate dust absorption seen in their composite SED.
Nevertheless, this is inconclusive because $N_{\rm H}$ cannot be constrained for the eFEDS--W4-nonX subsample given the current dataset.

\begin{table}
\caption{Absorption-corrected X-ray luminosities in the 2--10 keV band of the stacked eFEDS--W4-nonX subsample.}
\label{T_stack_Lx}
\centering
\begin{tabular}{lrr}
\hline \hline
redshift		& 	Assumed $N_{\rm H}$	($10^{22}$ cm$^{-2}$)	&	$L_{\rm X}$ (2--10 keV) \\
\hline			
\multirow{4}{*}{0.05} 	&	0		&	$2.95 \times 10^{40}$			\\
						&	1 		&	$1.40 \times 10^{41}$			\\
 						&	10 		&	$< 1.83 \times 10^{42}$\tablefootmark{a}			\\
 						&	100 	&	$2.98 \times 10^{42}$			\\
\hline						
\multirow{4}{*}{0.175} 	&	0		&	$2.81 \times 10^{41}$			\\
						&	1 		&	$1.03 \times 10^{42}$			\\
 						&	10 		&	$1.85 \times 10^{43}$			\\
 						&	100 	&	$2.83 \times 10^{43}$			\\											
\hline		
\multirow{4}{*}{0.375} 	&	0		&	$9.95 \times 10^{41}$			\\
						&	1 		&	$2.69 \times 10^{42}$			\\
 						&	10 		&	$3.86 \times 10^{43}$			\\
 						&	100 	&	$1.00 \times 10^{44}$			\\						
\hline	
\multirow{4}{*}{0.75} 	&	0		&	$4.57 \times 10^{42}$			\\
						&	1 		&	$8.53 \times 10^{42}$			\\
 						&	10 		&	$7.01 \times 10^{43}$			\\
 						&	100 	&	$4.60 \times 10^{44}$			\\		
\hline		
\multirow{4}{*}{2.0} 	&	0		&	$8.41 \times 10^{43}$			\\
						&	1 		&	$1.01 \times 10^{44}$			\\
 						&	10 		&	$2.79 \times 10^{44}$			\\
 						&	100 	&	$4.00 \times 10^{45}$			\\	
\hline															
\end{tabular}
\tablefoot{
\tablefoottext{a}{3$\sigma$ upper limit.}
}
\end{table}

\section{Discussion}
\label{D}

\subsection{ Relationship between MIR and X-ray luminosities}
\label{D_L6Lx}

\begin{figure}
\centering
\includegraphics[width=0.45\textwidth]{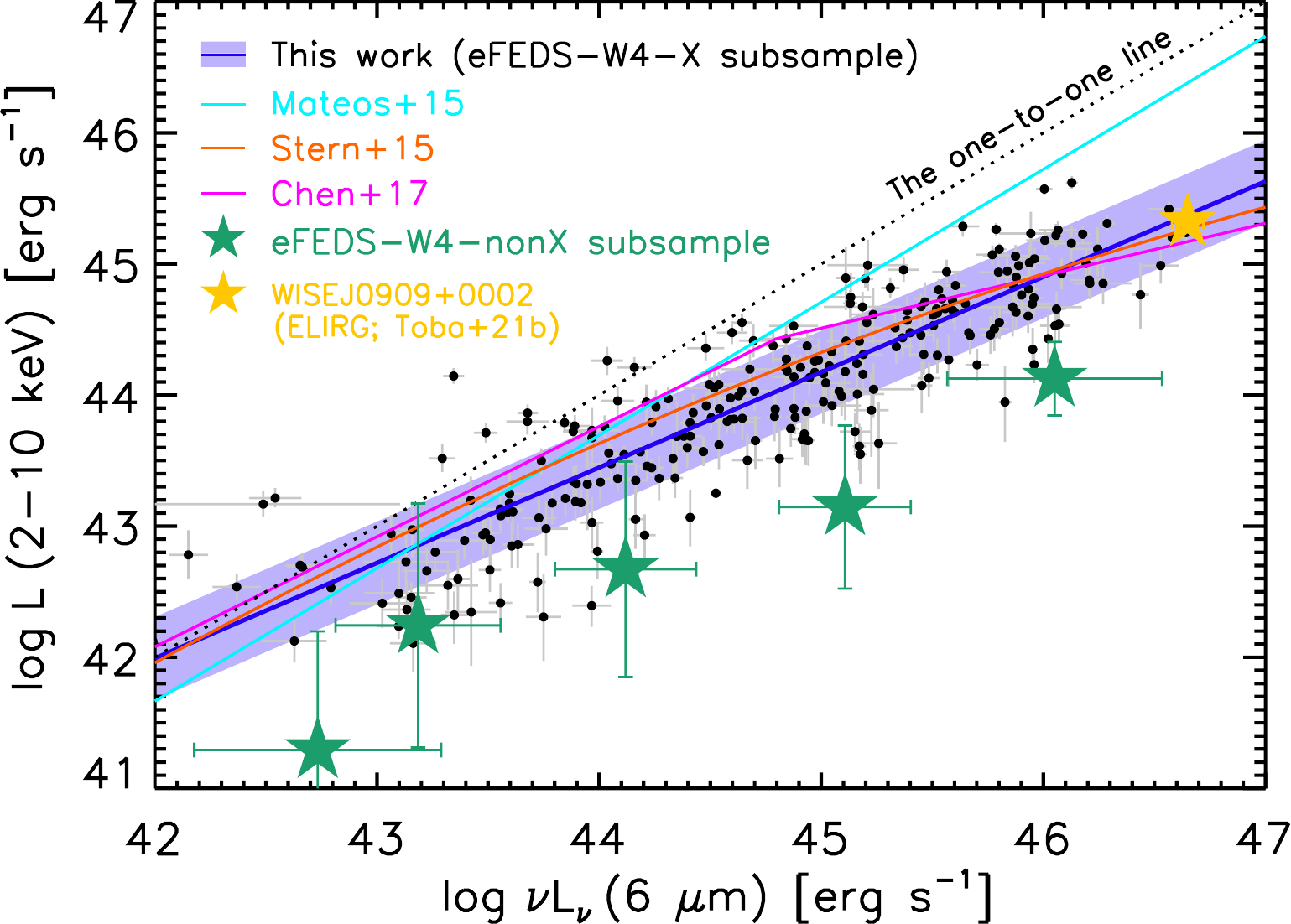}
\caption{Relationship between the rest-frame 6 $\mu$m luminosity contributed by AGN and the absorption-corrected, rest-frame 2--10-keV luminosity for the eFEDS--W4 sample. The best-fit linear function in log--log space and its 1$\sigma$ dispersion are shown as the solid blue line and the shaded blue band, respectively. The cyan line represents the best-fit for an X-ray-selected AGN sample from the BUXS catalog \citep{Mateos}. The orange and magenta lines show two-dimensional polynomial and bilinear relations from \citet{Stern} and \citet{Chen}, respectively. The green stars represent the eFEDS--W4-nonX subsample used for the X-ray stacking analysis in which we obtained the weighted mean and standard deviation in each redshift bin by taking into account the uncertainties in $N_{\rm H}$ ($N_{\rm H}$ = 0--10$^{23}$ cm$^{-2}$). The black dotted line represents a one-to-one correspondence between the 6 $\mu$m and 2--10 keV luminosities. The yellow star denotes the ELIRG (WISEJ0909+0002) reported in \cite{Toba_21b}.}
\label{fig_L6Lx}
\end{figure}

The MIR luminosity (e.g., the 6 $\mu$m luminosity $L_6$) and the hard X-ray luminosity [e.g., $L_{\rm X}$ (2--10 keV)] of the AGN are positively correlated over a wide luminosity range, regardless of the AGN type \citep[e.g.,][]{Gandhi,Asmus,Mateos,Stern,Chen,Ichikawa_17,Toba_19a}.
Because our AGN sample has a wide range of X-ray and IR luminosities, as shown in Figs.~\ref{fig_z_Lx_h} and \ref{fig_z_LIR}, it is worth investigating whether our eFEDS--W4-X sample also follows this relationship.
We show the resulting relationship between the rest-frame 6 $\mu$m luminosity and the rest-frame 2--10 keV luminosity in Fig.~\ref{fig_L6Lx}.
The plotted points represent X-ray AGN (with $\log\,L_{\rm X}$ (2--10 keV) $>$ 42 erg s$^{-1}$) with reliable $L_{\rm X}$ (i.e., which are classified as {\tt unobscured}, {\tt mildly-measured}, or {\tt well-measured}.
The X-ray luminosity is corrected for absorption (see Sect.~\ref{Xana}), while the 6 $\mu$m luminosity is corrected for contamination from the host galaxy in the same manner as in \cite{Toba_19a}.

We also plot the best-fit $L_{\rm 6}$--$L_{\rm X}$ relation for AGN samples taken from the Bright Ultra-hard XMM--Newton Survey (BUXS) \citep{Mateos}, the SDSS quasars \citep{Stern}, and data compiled for some deep fields \citep{Chen}. 
We performed a linear regression in log--log space for the eFEDS--W4-X AGN sample by considering the errors in $L_{\rm 6}$ and $L_{\rm X}$ obtained from a Bayesian maximum-likelihood method provided by \cite{Kelly}.
The resulting linear relation is as follows:
\begin{equation}
\label{Eq1}
\log L_{\rm X} = (0.73 \pm 0.02) \log L_{\rm 6} + (11.4 \pm 1.04),
\end{equation}
and the correlation coefficient is $r\sim0.91$, which confirms the tight correlation between $L_{6}$ and $L_{X}$ for our X-ray AGN sample.
We note that the slope of this relation (i.e., $L_{\rm X}/L_6$) is flatter than for the X-ray AGN detected by the {\it ROSAT} all-sky survey (RASS) reported in \citep{Toba_19a}, while it is steeper than that of the AGN detected by deep X-ray observations such as XMM--COSMOS \citep{Chen}.
\cite{Chen} suggested that X-ray flux limits may  affect the slopes, as the ratio $L_{\rm X}/L_6$ determined using shallower X-ray flux limits tends to be larger.
Because the flux limit of {\it eROSITA}/eFEDS is more than one order of magnitude deeper than that of RASS, while it is about one order of magnitude shallower than that of XMM--COSMOS (although the survey area of eFEDS is about 75 times larger than XMM--COSMOS), this result may support the explanation suggested by \cite{Chen}.

We also plot the 6 $\mu$m and hard X-ray luminosities of the eFEDS--W4-nonX stacked sample.
The mean values of $L_{\rm X}$ for this stacked sample (with $N_{\rm H}$ = 0--10$^{23}$ cm$^{-2}$) in each redshift bin, and their standard deviations, are plotted on the y-axis.
The weighted means and standard deviations of the $L_{\rm 6}$ values in each redshift bin are plotted on the x-axis.
We find that the best-fit slope of the stacked AGN sample with $\log\,L_{\rm X} > 42$ erg s$^{-1}$ is $\sim0.47$, which is consistent with that for the XMM--COSMOS sample, which further supports the explanation suggested by \cite{Chen}.

\subsection{Structural properties of AGN host galaxies}
\label{D_decomp}

In Sect.~\ref{s_2D}, we described the 2D image decomposition of the HSC images used to distinguish emission from a central point source and an extended component.
This analysis provides flux-density data for the AGN and its host separately, which enables us to estimate the stellar mass reliably without it being affected by the AGN emission.

\subsubsection{Correlation of the AGN fraction derived from {\tt X-CIGALE} and 2D image decomposition}

\begin{figure}
\centering
\includegraphics[width=0.45\textwidth]{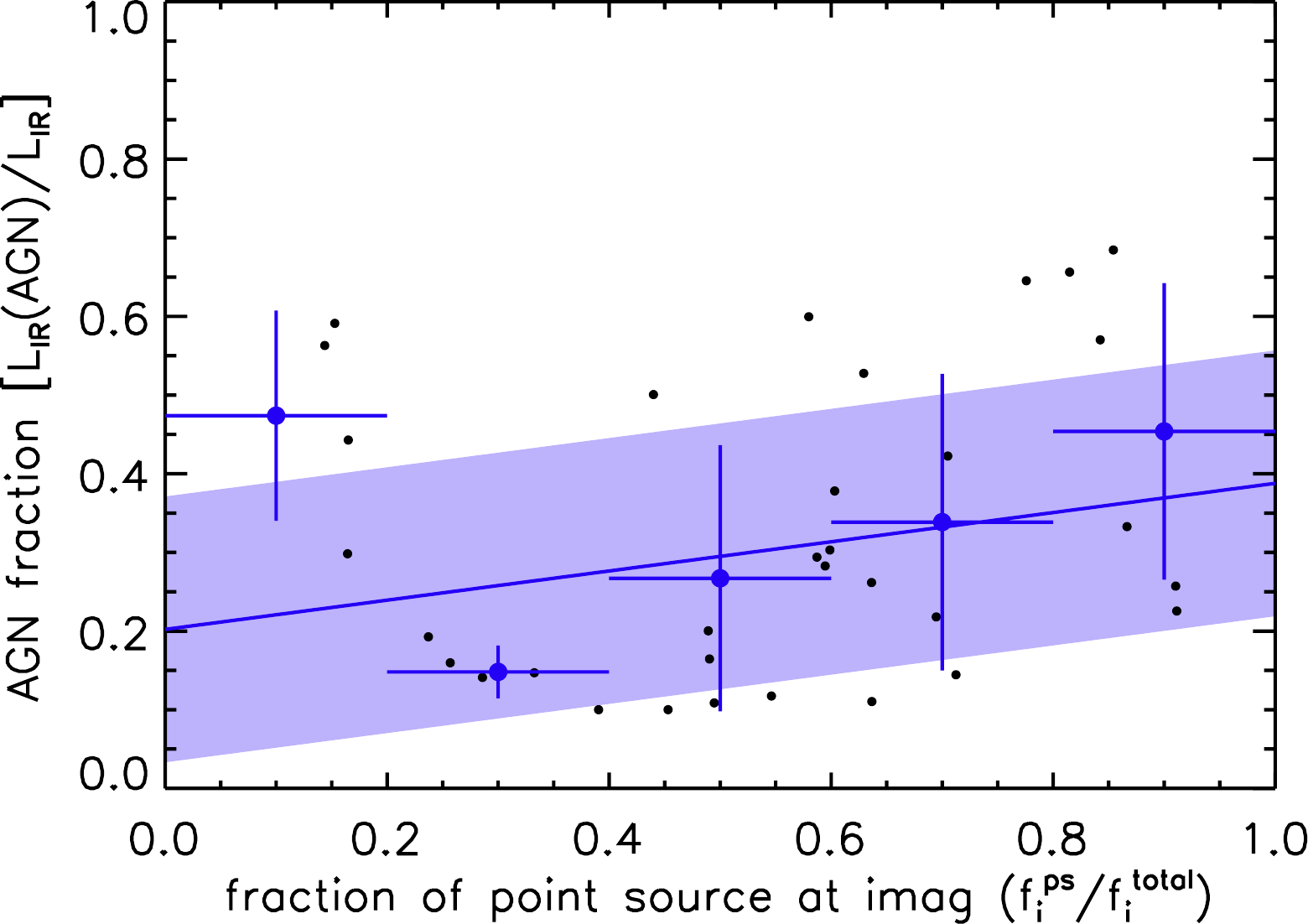}
\caption{AGN fraction ($L_{\rm IR}$(AGN)/$L_{\rm IR}$) of the eFEDS--W4-X sample at $0.2 < z < 0.8$ versus the point-source fraction of the $i$-band flux density ($f_{\rm i}^{\rm PS}$/$f_{\rm i}^{\rm total}$). The black points represent the data. The blue dots with error bars represent the mean and standard deviation in each bin for the eFEDS--W4-X sources classified as X-ray type 1 AGN. The solid line and shaded region denote the best-fit relation and the 1$\sigma$ uncertainty.}
\label{fig_f_AGN}
\end{figure}

Fig.~\ref{fig_f_AGN} shows the relationship between the point-source fraction of the $i$-band flux density  ($f_{\rm i}^{\rm PS}$/$f_{\rm i}^{\rm total}$) derived from the image decomposition and the AGN fraction derived from the SED fitting with {\tt X-CIGALE}.
The AGN fraction is defined as $L_{\rm IR}$(AGN)/$L_{\rm IR}$.
The sources plotted in this figure are limited to those with reduced $\chi_{\rm decomp}^2 < 5$ at $0.2 < z < 0.8$ (see Sect. \ref{s_2D}).
In particular, it is worth investigating how the AGN fraction from an optical decomposed-image point of view is associated with that from the IR SED-fitting point of view for type 1 AGN.
Hence, we focus on X-ray type 1 AGN, as classified in Sect.~\ref{R_X}.
We find these quantities to be weakly correlated, with correlation coefficients $r\sim 0.26$ where $r$ is derived using the Bayesian method \citep{Kelly} (see Sect.~\ref{D_L6Lx}).
This result indicates that the 2D image decomposition and SED fitting with {\tt X-CIGALE} work consistently to evaluate the contributions of the AGN to the optical and IR bands.

In addition, we find that even for objects with $f_{\rm i}^{\rm PS}$/$f_{\rm i}^{\rm total} > 0.8$ the AGN fraction rarely exceeds 0.6, which is consistent with what is reported in previous works \citep[e.g.,][]{Lyu,Ichikawa_19}.
The SF activity is expected to contribute significantly to the IR fluxes for luminous AGNs, suggesting an AGN--SF connection; i.e., the co-existence of gas consumption into SF and supermassive black holes (SMBHs) \citep[see e.g.,][]{Imanishi,Stemo}.

Further, we compare the AGN fraction measured at the rest-frame $i$-band from the best-fit SEDs and image decomposition ($f_{\rm i}^{\rm PS}$/$f_{\rm i}^{\rm total}$).
We observe a weak correlation with correlation coefficient $r\sim 0.2$ for X-ray type 1 AGN.
By contrast, we observe a moderately strong correlation ($r\sim 0.3$) for X-ray type 2 AGN, which may indicate that SED decomposition (to distinguish between emissions from the accretion disk and stellar component) in the optical bands has a large uncertainty, particularly for type 1 AGN (see Sect.~\ref{D_Mhost}).

\subsubsection{Consistency of stellar mass based on the host flux densities}
\label{D_Mhost}

Many works have reported that estimates of stellar masses based on SED fitting---particularly for type 1 AGN---may have large uncertainties because it is often difficult to distinguish the accretion disk and the stellar component given a limited number of photometric bands \citep[see e.g.,][]{Merloni10,Bongiorno,Toba_18}.
This motivated an investigation to determine how to estimate securely the stellar mass derived from SED fitting with {\tt X-CIGALE}.
The host flux information provided by the 2D image decomposition enables us to estimate the stellar mass reliably \citep[e.g.,][]{Li}.
Accordingly, we performed SED fitting to the HSC five-band photometry of each AGN host to obtain the stellar mass ($M_*^{\rm HSC}$) without being affected by AGN emission.
To obtain a better estimate of the stellar mass, we parameterized the SFH and SSP over narrow intervals.

\begin{figure}
\centering
\includegraphics[width=0.48\textwidth]{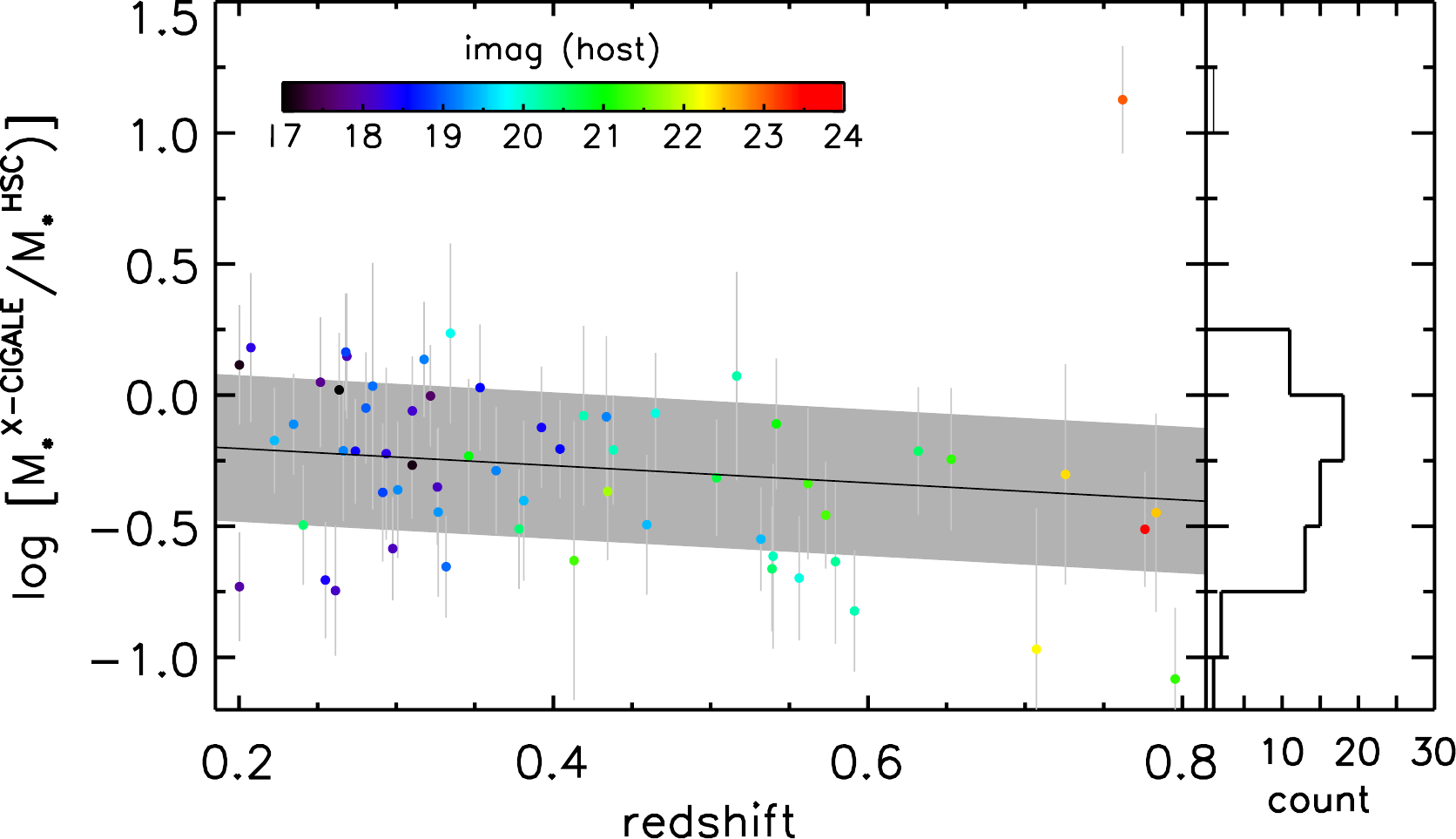}
\caption{Ratio of the stellar mass derived from SED fitting with all the data ($M_*^{\rm X-CIGALE}$) to that obtained from the HSC five-band data for the AGN host galaxy ($M_*^{\rm HSC}$) as a function of redshift, color-coded by the $i$-band magnitude of the AGN host. The black line and shaded region represent the best-fit relationship and the 1$\sigma$ uncertainty, respectively. The histogram of the ratio is shown in the right panel.}
\label{fig_M}
\end{figure}

Fig.~\ref{fig_M} shows a comparison of the stellar mass derived from the SED fitting with all the data ($M_*^{\rm X-CIGALE}$; see Sect.~\ref{s_SED}) and from the HSC five-band flux density for the AGN host from the eFEDS--W4-X sample at $0.2 < z < 0.8$ as a function of redshift.
This comparison shows that the stellar mass derived in this work may be systematically underestimated by $\sim$0.3 dex.
The resulting correlation coefficient ($r = -0.20 \pm 0.17$) indicates that this offset does not considerably depend on redshift, at least up to $z = 0.8$.
Nevertheless, a faint ($i$-mag $>$ 23) object at $z > 0.7$ deviates considerably from the typical offset, which should be kept in mind as indicative of the uncertainty in the stellar mass.

\subsection{Evolutionary stage of the eROSITA-detected MIR galaxies}
\label{D_NH_EDD}

Finally, we discuss the evolutionary phase of the eFEDS--W4 sample using the Eddington ratio ($\lambda_{\rm Edd}$) and $N_{\rm H}$.
The Eddington ratio is defined as $\lambda_{\rm Edd} = L_{\rm bol}/L_{\rm Edd}$, where $L_{\rm bol}$ is the bolometric luminosity and $L_{\rm Edd}$ is the Eddington luminosity.
For 32 quasars that are cataloged in the SDSS DR16 quasar catalog \citep{Lyke}, we estimate the single-epoch virial black-hole mass ($M_{\rm BH}$) based on a few recipes using the continuum luminosities at 1450, 3000, and 5100 \AA, as well as the FWHM of the C{\,\sc iv},  Mg{\,\sc ii}, and H$\beta$ lines \citep{Vestergaard,Shen} \citep[see also Sect.~5.2 in][]{Rakshit}.
We employed the quasar spectral-fitting package \citep[{\tt QSFit} v1.3.0;][]{Calderone} to determine the line widths and continuum luminosities in the same manner as in \cite{Toba_21a}.
The values of $M_{\rm BH}$ for the remaining sources are obtained from the stellar mass using an empirical relation for AGN up to $z\sim 2.5$ provided in \cite{Suh}.  
We use $M_*^{\rm X-CIGALE}$ as the stellar mass unless $M_*^{\rm HSC}$ is available.
The uncertainties in the stellar mass and the intrinsic scatter of $M_*$--$M_{\rm BH}$ ($\sim0.5$ dex) reported in \cite{Suh} are propagated to the uncertainty of the converted $M_{\rm BH}$.
The value of $L_{\rm bol}$ is estimated by integrating the best-fit SED template for the AGN component output from {\tt X-CIGALE} \citep[see e.g.,][]{Toba_17c}.
The uncertainty in $L_{\rm bol}$ is calculated by error propagation of the uncertainty in the AGN luminosity.

\begin{figure}
\centering
\includegraphics[width=0.48\textwidth]{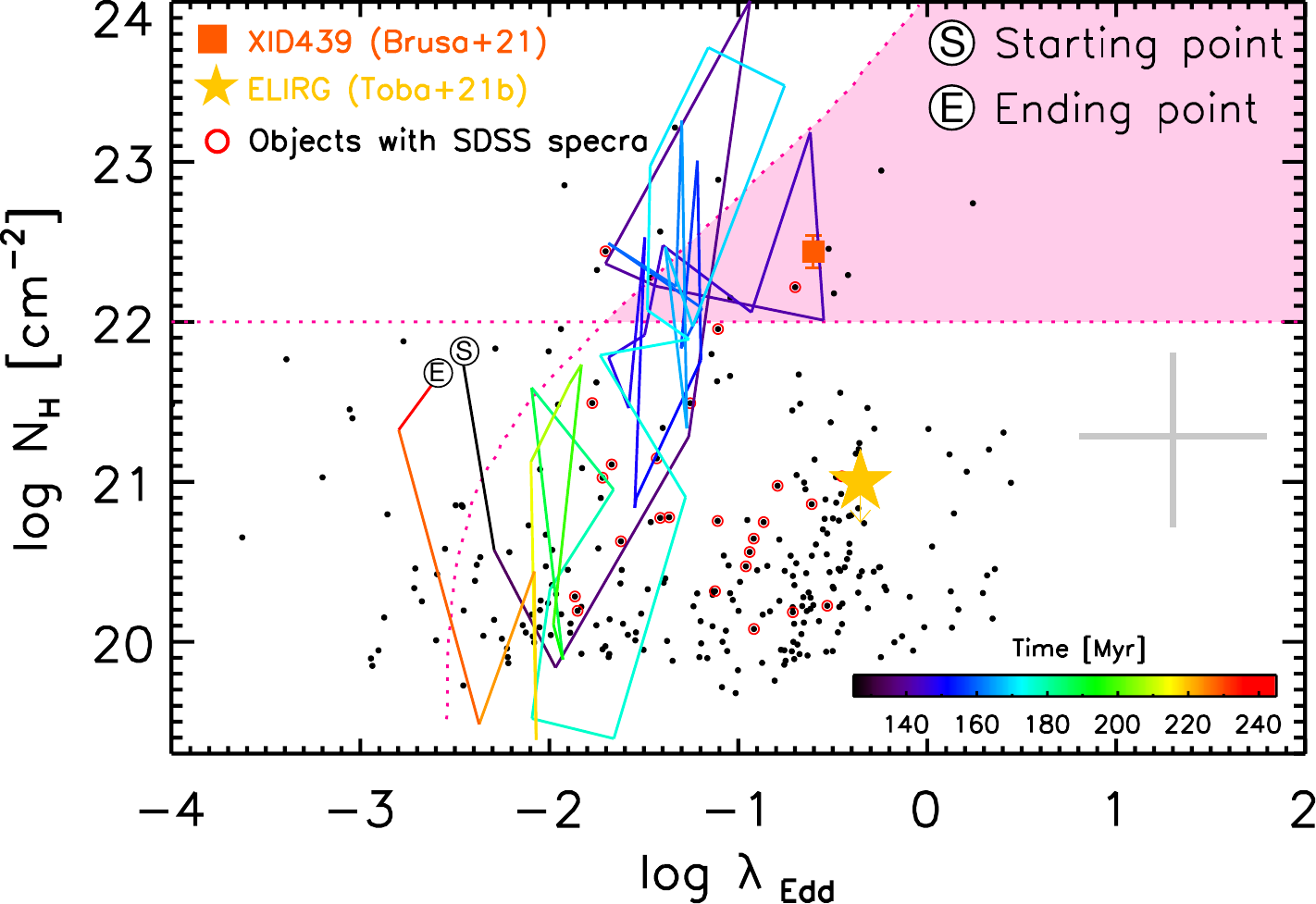}
\caption{Relation between $\log\, N_{\rm H}$ and $\lambda_{\rm Edd}$ for the eFEDS--W4 sample. The yellow star represents the ELIRG \citep{Toba_21b}, and the red square shows the location of XID439 (see text). Objects with red circles have optical spectra from SDSS DR16. Typical uncertainties in $\log\, N_{\rm H}$ and $\lambda_{\rm Edd}$ are shown at the bottom right. The dotted line denotes the effective Eddington limit ($\lambda_{\rm Edd}^{\rm eff}$; \citealt{Fabian_06}) for different values of $N_{\rm H}$ \citep{Fabian,Ricci}. The pink shaded area represents the blow-out region, in which the dusty gas that produces the nuclear obscuration is expected to be pushed away by radiation pressure (i.e., $\lambda_{\rm Edd} > \lambda_{\rm Edd}^{\rm eff}$). The solid line shows the evolutionary track calculated by \cite{Yutani}, color-coded by time in units of Myr.}
\label{fig_NH_Edd}
\end{figure}

\begin{figure}
\centering
\includegraphics[width=0.45\textwidth]{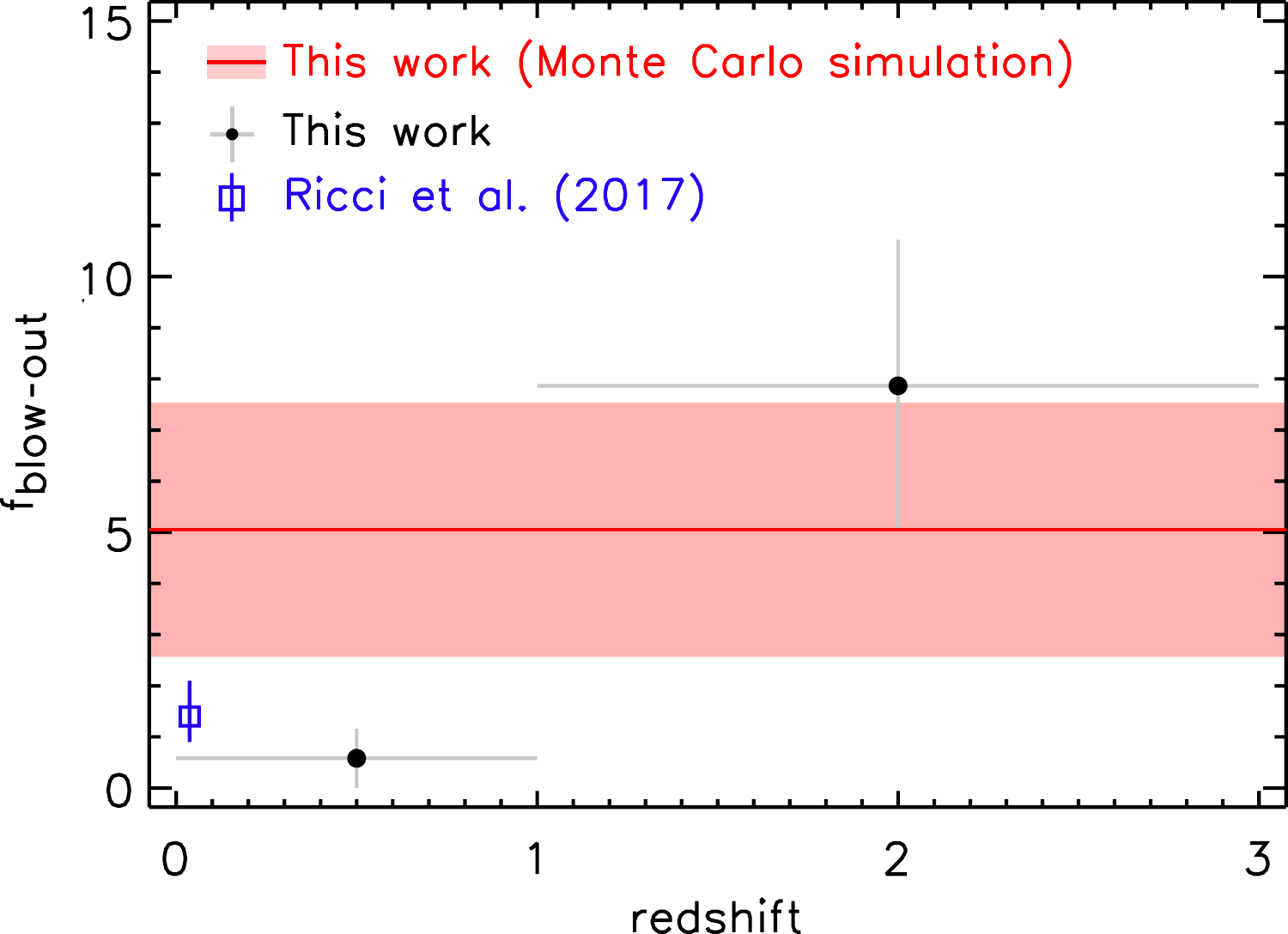}
\caption{Fraction of objects in the blow-out region ($f_{\rm blow-out}$) as a function of redshift. The blue square denotes the hard-X-ray-selected AGN sample \citep{Ricci}, while the black points are our eFEDS--W4-X sample. The red line and shaded region represent $f_{\rm blow-out}$ calculated from the Monte Carlo simulation.}
\label{fig_z_fBO}
\end{figure}

Fig~\ref{fig_NH_Edd} shows the relation between $\lambda_{\rm Edd}$ and $N_{\rm H}$ for the eFEDS--W4-X subsample, which tells us how the obscuration of the AGN is associated with mass accretion onto the SMBH \citep[see e.g.,][]{Fabian}.
Objects in the pink shaded region in Fig~\ref{fig_NH_Edd} are expected to blow out the surrounding gas and dust by strong radiation pressure \citep{Fabian_06,Fabian,Ishibashi}.
We estimate the fraction of objects located in this ``blow-out region'' ($f_{\rm blow-out}$).
Note that obtained values of $N_{\rm H}$ and $\lambda_{\rm Edd}$ have quite large uncertainties (0.5--1 dex).
In addition, only 12\% of the eFEDS--W4-X subsample plotted in Fig.~\ref{fig_NH_Edd} has a spectroscopically derived $\lambda_{\rm Edd}$, and $\lambda_{\rm Edd}$ for the majority of the sample is empirically estimated by assuming an $M_*$--$M_{\rm BH}$ relation.
Therefore, we take into account the uncertainties in $N_{\rm H}$ and $\lambda_{\rm Edd}$ as well as statistical errors in the fraction through Monte Carlo randomization to estimate $f_{\rm blow-out}$.
We find that $f_{\rm blow-out} = 5.0 \pm 2.5$ \% (see Fig.~\ref{fig_z_fBO}). 
We note that this value should be considered a lower limit because we only investigated $N_{\rm H}$--$\lambda_{\rm Edd}$ for X-ray detected {\it WISE} 22 $\mu$m sources with $z_{\rm spec}$ (i.e., eFEDS--W4-X subsample).
In particular, a fraction of X-ray undetected {\it WISE} 22 $\mu$m sources (i.e., eFEDS--W4-nonX sample) is expected to be obscured AGN (see Sect.~\ref{R_stack}); thus, these AGN may be located in the blow-out region, which would increase $f_{\rm blow-out}$.

Nevertheless, the estimated $f_{\rm blow-out}$ is larger than that reported by \cite{Ricci}, who examined the $\lambda_{\rm Edd}$--$N_{\rm H}$ relation\footnote{The values of $N_{\rm H}$ and $\lambda_{\rm Edd}$ were estimated by \cite{Ricci} using X-ray and optical spectra. Hence, those values are well constrained compared with this work.} for hard-X-ray-selected AGN with a median redshift of $z = 0.037$ and reported that only 1.4\% of the sources in their sample are found in this region.
This discrepancy is probably caused by the difference in redshift; the fraction of X-ray-obscured AGN increases with increasing redshift \citep[e.g.,][]{Hasinger,Merloni_14,Ueda}.
Fig.~\ref{fig_z_fBO} shows the fraction of objects in the blow-out region as a function of redshift.
There is a positive correlation: high-$z$ obscured galaxies tend to reside in the AGN-feedback phase, which supports the above possibility.
Recently, \cite{Jun} investigated the $\lambda_{\rm Edd}$--$N_{\rm H}$ relationship for optical/IR/submillimeter-selected AGN as well as for X-ray selected ones with redshifts up to $z\sim 3$.
They found that IR/submillimeter-bright red AGN tend to be located in the blow-out region.
We note that the absence of local AGN in the blow-out region may also be coupled with the timescale of the blow-out phase \citep[see][for details]{Jun}.
Given the fact that (i) the $M_*$--$M_{\rm BH}$ relation used to derive $\lambda_{\rm Edd}$ may be applicable up to $z = 2.5$ \citep{Suh} and (ii) the number of objects in this redshift range is too small (see Fig.~\ref{fig_zhist}), we may need more sample spectra to determine whether this decreasing trend is real.

\cite{Yutani} performed high-resolution N-body/SPH simulations with the {\tt ASURA} code \citep{Saitoh_08,Saitoh_09} and investigated the time evolution of the SEDs of galaxy mergers using the radiative transfer simulation code {\tt RADMC-3D} \citep{Dullemond}.
Although our eFEDS--W4-X sources may not always experience galaxy mergers in their lifetimes, it is still worth comparing observational results with model predictions.
The evolutionary track of a merger as a function of time on the $\lambda_{\rm Edd}$--$N_{\rm H}$ diagram is presented in Fig.~\ref{fig_NH_Edd}.
This track is calculated from the final ($\sim$ 200 Myr) phase of evolution of the merger of two galactic central regions, each with $M_{\rm BH} = 4 \times 10^7$ $M_{\sun}$.
The stellar mass and the gas mass of the galactic core are $M_* = 4 \times 10^9$ $M_{\sun}$ and $M_g = 8\times 10^8$  $M_{\sun}$, respectively.
The spatial resolution is 8 pc.
The gas-to-dust mass ratio is assumed to be 50 \citep{Toba_17d}.
We observe that approximately half of our eFEDS--W4--X subsample, including ELIRG, is not covered by evolutionary tracks.
This is probably because the assumed $M_{\rm BH}$ in \cite{Yutani} is smaller than that of objects considered in this work.
The typical $M_{\rm BH}$ for objects with spectroscopically derived $M_{\rm BH}$ is $\log\, (M_{\rm BH}/M_{\sun}) \sim 9.3$.
Indeed, \cite{Yutani} could not reproduce objects with $L_{\rm IR} > 10^{13}$ $L_{\sun}$ under the condition described above, which may indicate that more massive BH mergers are required to reproduce the $\lambda_{\rm Edd}$--$N_{\rm H}$ distribution for our sample.
Additionally, we find that the track moves around significantly on the $\lambda_{\rm Edd}$--$N_{\rm H}$ plot. In particular, $\lambda_{\rm Edd}$ changes dramatically by 0.5 or even 1 dex in the blow-out region.
The timescale for the blow-out/feedback phase is expected to be $\sim30$ Myr, which is consistent with that reported in \cite{Jun}. 

Actually, \cite{Brusa} reported a powerful quasar (XID439) with red colors in the optical, MIR, and X-ray at $z = 0.603$ detected by {\it eROSITA}. 
Its SDSS spectrum shows a broad (FWHM $\sim$1650 km s$^{-1}$) component in the [O\,{\sc iii}]$\lambda$5007 line, suggesting that XID439 is in the AGN-feedback phase.
Multiwavelength SED fitting with {\tt X-CIGALE} and X-ray spectral analysis yielded the values $\lambda_{\rm Edd} \sim 0.25$ and $N_{\rm H} = 2.75 \times 10^{22}$ cm$^{-2}$ for the quasar.
XID439 is thus in the blow-out region and on the evolutionary track in the feedback phase in $\lambda_{\rm Edd}$--$N_{\rm H}$ plane (see Fig.~\ref{fig_NH_Edd}), which supports the above idea.
As mentioned before, however, the relatively small number of objects with spectroscopically derived values of $\lambda_{\rm Edd}$ still prevents us from drawing a definite conclusion about this argument.
Our ongoing spectroscopic campaign through the SDSS IV/V collaboration will provide a more complete view of this figure.

\section{Summary}
\label{Sum}

In this work, we have investigated the physical properties of MIR galaxies at $z <4$.
The parent sample is drawn from 7,780 {\it WISE} 22 $\mu$m (W4)-detected sources (the eFEDS--W4 sample) in the eFEDS catalog within a region of 140 deg$^2$.
By cross-matching the sample with the eFEDS main X-ray catalog, we find that 692 MIR galaxies (the eFEDS--W4-X sample) are detected and 7,088 sources (the eFEDS--W4-nonX sample) are undetected by {\it eROSITA}.
We compiled multiwavelength data from X-ray to FIR wavelengths and performed SED fitting with {\tt X-CIGALE}.
We performed X-ray spectral analysis for the eFEDS--W4 sample and X-ray stacking analysis for the eFEDS--W4-nonX sample.
This multiwavelength approach provides AGN and host properties (such as the stellar mass, SFR, and $N_{\rm H}$) for the eFEDS--W4 galaxies.

Because of the wide-area data from eFEDS, we are able to determine various physical properties for the sample objects and find candidates of spatially rare populations such as ELIRGs.
With all the caveats discussed in Sect~\ref{D_NH_EDD} in mind, the distribution of the eFEDS--W4 sample on the $\lambda_{\rm Edd}$--$N_{\rm H}$ plane, and a comparison with the output from a galaxy-merger simulation, indicate that approximately 5.0\% of the sources in our sample are likely to be in a relatively short-lived ($\sim$30 Myr) feedback phase, in which nuclear material is blown out owing to radiation pressure from the AGN.

The {\it eROSITA} all-sky survey (eRASS) is ongoing and is planned to continue until the end of 2023.
Because SRG/eROSITA observes the entire sky in six months and will scan the entire sky eight times by the end of the mission, depth of the survey will increase progressively.
The completion of eRASS (eRASS8) will thus provide a complete census of MIR galaxies over a wide luminosity and redshift range from the X-ray point of view.
In that sense, this work establishes a benchmark for obtaining a full picture of such IR populations.

\begin{acknowledgements}

We thank the anonymous referee for the insightful comments and suggestions that improved the paper.
We gratefully acknowledge Dr. Andy Goulding for useful discussion and comments,
We also thank Dr. Claudio Ricci for helping to make Figure \ref{fig_NH_Edd}.\\

This work is based on data from {\it eROSITA}, the primary instrument aboard SRG, a joint Russian-German science mission supported by the Russian Space Agency (Roskosmos), in the interests of the Russian Academy of Sciences represented by its Space Research Institute (IKI), and the Deutsches Zentrum f\"ur Luft- und Raumfahrt (DLR). 
The SRG spacecraft was built by Lavochkin Association (NPOL) and its subcontractors, and is operated by NPOL with support from the Max-Planck Institute for Extraterrestrial Physics (MPE).
The development and construction of the {\it eROSITA} X-ray instrument was led by MPE, with contributions from the Dr. Karl Remeis Observatory Bamberg \& ECAP (FAU Erlangen-Nuernberg), the University of Hamburg Observatory, the Leibniz Institute for Astrophysics Potsdam (AIP), and the Institute for Astronomy and Astrophysics of the University of T\"ubingen, with the support of DLR and the Max Planck Society. 
The Argelander Institute for Astronomy of the University of Bonn and the Ludwig Maximilians Universit\"at Munich also participated in the science preparation for {\it eROSITA}. 
The {\it eROSITA} data shown here were processed using the eSASS/NRTA software system developed by the German {\it eROSITA} consortium. \\

The Hyper Suprime-Cam (HSC) collaboration includes the astronomical communities of Japan and Taiwan, and Princeton University.  The HSC instrumentation and software were developed by the National Astronomical Observatory of Japan (NAOJ), the Kavli Institute for the Physics and Mathematics of the Universe (Kavli IPMU), the University of Tokyo, the High Energy Accelerator Research Organization (KEK), the Academia Sinica Institute for Astronomy and Astrophysics in Taiwan (ASIAA), and Princeton University.  Funding was contributed by the FIRST program from the Japanese Cabinet Office, the Ministry of Education, Culture, Sports, Science and Technology (MEXT), the Japan Society for the Promotion of Science (JSPS), Japan Science and Technology Agency  (JST), the Toray Science  Foundation, NAOJ, Kavli IPMU, KEK, ASIAA, and Princeton University.
This paper makes use of software developed for the Large Synoptic Survey Telescope. We thank the LSST Project for making their code available as free software at  http://dm.lsst.org 
This paper is based [in part] on data collected at the Subaru Telescope and retrieved from the HSC data archive system, which is operated by Subaru Telescope and Astronomy Data Center (ADC) at NAOJ. Data analysis was in part carried out with the cooperation of Center for Computational Astrophysics (CfCA), NAOJ.\\
The Pan-STARRS1 Surveys (PS1) and the PS1 public science archive have been made possible through contributions by the Institute for Astronomy, the University of Hawaii, the Pan-STARRS Project Office, the Max Planck Society and its participating institutes, the Max Planck Institute for Astronomy, Heidelberg, and the Max Planck Institute for Extraterrestrial Physics, Garching, The Johns Hopkins University, Durham University, the University of Edinburgh, the Queen’s University Belfast, the Harvard-Smithsonian Center for Astrophysics, the Las Cumbres Observatory Global Telescope Network Incorporated, the National Central University of Taiwan, the Space Telescope Science Institute, the National Aeronautics and Space Administration under grant No. NNX08AR22G issued through the Planetary Science Division of the NASA Science Mission Directorate, the National Science Foundation grant No. AST-1238877, the University of Maryland, Eotvos Lorand University (ELTE), the Los Alamos National Laboratory, and the Gordon and Betty Moore Foundation.\\

The Legacy Surveys consist of three individual and complementary projects: the Dark Energy Camera Legacy Survey (DECaLS; Proposal ID \#2014B-0404; PIs: David Schlegel and Arjun Dey), the Beijing-Arizona Sky Survey (BASS; NOAO Prop. ID \#2015A-0801; PIs: Zhou Xu and Xiaohui Fan), and the Mayall z-band Legacy Survey (MzLS; Prop. ID \#2016A-0453; PI: Arjun Dey). DECaLS, BASS and MzLS together include data obtained, respectively, at the Blanco telescope, Cerro Tololo Inter-American Observatory, NSF's NOIRLab; the Bok telescope, Steward Observatory, University of Arizona; and the Mayall telescope, Kitt Peak National Observatory, NOIRLab. The Legacy Surveys project is honored to be permitted to conduct astronomical research on Iolkam Du'ag (Kitt Peak), a mountain with particular significance to the Tohono O'odham Nation.

NOIRLab is operated by the Association of Universities for Research in Astronomy (AURA) under a cooperative agreement with the National Science Foundation.

This project used data obtained with the Dark Energy Camera (DECam), which was constructed by the Dark Energy Survey (DES) collaboration. Funding for the DES Projects has been provided by the U.S. Department of Energy, the U.S. National Science Foundation, the Ministry of Science and Education of Spain, the Science and Technology Facilities Council of the United Kingdom, the Higher Education Funding Council for England, the National Center for Supercomputing Applications at the University of Illinois at Urbana-Champaign, the Kavli Institute of Cosmological Physics at the University of Chicago, Center for Cosmology and Astro-Particle Physics at the Ohio State University, the Mitchell Institute for Fundamental Physics and Astronomy at Texas A\&M University, Financiadora de Estudos e Projetos, Fundacao Carlos Chagas Filho de Amparo, Financiadora de Estudos e Projetos, Fundacao Carlos Chagas Filho de Amparo a Pesquisa do Estado do Rio de Janeiro, Conselho Nacional de Desenvolvimento Cientifico e Tecnologico and the Ministerio da Ciencia, Tecnologia e Inovacao, the Deutsche Forschungsgemeinschaft and the Collaborating Institutions in the Dark Energy Survey. The Collaborating Institutions are Argonne National Laboratory, the University of California at Santa Cruz, the University of Cambridge, Centro de Investigaciones Energeticas, Medioambientales y Tecnologicas-Madrid, the University of Chicago, University College London, the DES-Brazil Consortium, the University of Edinburgh, the Eidgenossische Technische Hochschule (ETH) Zurich, Fermi National Accelerator Laboratory, the University of Illinois at Urbana-Champaign, the Institut de Ciencies de l'Espai (IEEC/CSIC), the Institut de Fisica d'Altes Energies, Lawrence Berkeley National Laboratory, the Ludwig Maximilians Universitat Munchen and the associated Excellence Cluster Universe, the University of Michigan, NSF's NOIRLab, the University of Nottingham, the Ohio State University, the University of Pennsylvania, the University of Portsmouth, SLAC National Accelerator Laboratory, Stanford University, the University of Sussex, and Texas A\&M University.

BASS is a key project of the Telescope Access Program (TAP), which has been funded by the National Astronomical Observatories of China, the Chinese Academy of Sciences (the Strategic Priority Research Program ``The Emergence of Cosmological Structures'' Grant \# XDB09000000), and the Special Fund for Astronomy from the Ministry of Finance. The BASS is also supported by the External Cooperation Program of Chinese Academy of Sciences (Grant \# 114A11KYSB20160057), and Chinese National Natural Science Foundation (Grant \# 11433005).

The Legacy Survey team makes use of data products from the Near-Earth Object Wide-field Infrared Survey Explorer (NEOWISE), which is a project of the Jet Propulsion Laboratory/California Institute of Technology. NEOWISE is funded by the National Aeronautics and Space Administration.

The Legacy Surveys imaging of the DESI footprint is supported by the Director, Office of Science, Office of High Energy Physics of the U.S. Department of Energy under Contract No. DE-AC02-05CH1123, by the National Energy Research Scientific Computing Center, a DOE Office of Science User Facility under the same contract; and by the U.S. National Science Foundation, Division of Astronomical Sciences under Contract No. AST-0950945 to NOAO.

Based on observations made with ESO Telescopes at the La Silla Paranal Observatory under programme IDs 177.A-3016, 177.A-3017, 177.A-3018 and 179.A-2004, and on data products produced by the KiDS consortium. The KiDS production team acknowledges support from: Deutsche Forschungsgemeinschaft, ERC, NOVA and NWO-M grants; Target; the University of Padova, and the University Federico II (Naples).\\

Funding for the Sloan Digital Sky Survey IV has been provided by the Alfred P. Sloan Foundation, the U.S. Department of Energy Office of Science, and the Participating Institutions. 

SDSS-IV acknowledges support and resources from the Center for High Performance Computing  at the University of Utah. The SDSS website is www.sdss.org.

SDSS-IV is managed by the Astrophysical Research Consortium for the Participating Institutions of the SDSS Collaboration including the Brazilian Participation Group, the Carnegie Institution for Science, Carnegie Mellon University, Center for Astrophysics | Harvard \& Smithsonian, the Chilean Participation Group, the French Participation Group, Instituto de Astrof\'isica de Canarias, The Johns Hopkins University, Kavli Institute for the 
Physics and Mathematics of the Universe (IPMU) / University of Tokyo, the Korean Participation Group, Lawrence Berkeley National Laboratory, Leibniz Institut f\"ur Astrophysik Potsdam (AIP),  Max-Planck-Institut f\"ur Astronomie (MPIA Heidelberg), Max-Planck-Institut f\"ur Astrophysik (MPA Garching), Max-Planck-Institut f\"ur 
Extraterrestrische Physik (MPE), National Astronomical Observatories of China, New Mexico State University, 
New York University, University of Notre Dame, Observat\'ario Nacional / MCTI, The Ohio State University, Pennsylvania State University, Shanghai Astronomical Observatory, United Kingdom Participation Group, Universidad Nacional Aut\'onoma de M\'exico, University of Arizona, University of Colorado Boulder, University of Oxford, University of Portsmouth, University of Utah, University of Virginia, University of Washington, University of 
Wisconsin, Vanderbilt University, and Yale University.\\

The {\it Herschel}-ATLAS is a project with Herschel, which is an ESA space observatory with science instruments provided by European-led Principal Investigator consortia and with important participation from NASA. The H-ATLAS website is \url{http://www.h-atlas.org/}.\\

This research is based on observations with AKARI, a JAXA project with the participation of ESA.\\

Numerical computations/simulations were carried out (in part) using the SuMIRe cluster operated by the Extragalactic
OIR group at ASIAA.\\

This work is supported by JSPS KAKENHI grant Nos. 18J01050, 19K14759, and 19KK0076 (Y.Toba), 20H01946 (Y.Ueda), 20H01949 (T.Nagao), and 20K04014 (Y.Terashima).

\end{acknowledgements}


\end{document}